\documentclass[a4paper,fleqn]{cas-dc}

\usepackage{natbib}
\setcitestyle{square,sort,comma,numbers,sort&compress}
\usepackage[linesnumbered,ruled]{algorithm2e}
\usepackage{url}
\usepackage{multirow}

\usepackage{adjustbox}
\usepackage{makecell}
\usepackage{enumitem}
\def\tsc#1{\csdef{#1}{\textsc{\lowercase{#1}}\xspace}}
\tsc{WGM}
\tsc{QE}

\begin{document}
\let\WriteBookmarks\relax
\def\floatpagepagefraction{1}
\def\textpagefraction{.001}

\shorttitle{Grid-Aware On-Route Fast-Charging Infrastructure Planning for BEB with Equity Considerations}    

\shortauthors{X. Zhao, C. Zhao, G. Jia}  

\title [mode = title]{Grid-Aware On-Route Fast-Charging Infrastructure Planning for Battery Electric Bus with Equity Considerations: A Case Study in South King County}

\author[1]{Xinyi Zhao}[orcid = 0000-0002-9655-4889]

\author[1]{Chaoyue Zhao}
\ead{cyzhao@uw.edu}
\cormark[1]

\author[2]{Grace Jia}

\affiliation[1]{organization={Department of Industrial \& Systems Engineering, University of Washington},
            city={Seattle},
            postcode={98195}, 
            state={Washington},
            country={USA}}

\affiliation[2]{organization={Department of Civil \& Environmental Engineering, University of Washington},
            city={Seattle},
            postcode={98195}, 
            state={Washington},
            country={USA}}

\cortext[1]{Corresponding author}

\begin{abstract}
The transition from traditional bus fleets to zero-emission ones necessitates the development of effective planning models for battery electric bus (BEB) charging infrastructure. On-route fast charging stations, distinct from on-base charging stations, present unique challenges related to safe operation and power supply capacity, making it difficult to control grid operational costs. This paper establishes a novel framework that integrates the bus route network and power network, which leverages the inter-dependency between both networks to optimize the planning outcomes of on-route BEB charging stations in South King County. The problem is formulated as a mixed-integer second-order cone programming model, aiming to minimize the overall planning cost, which includes investments in charging equipment, power facility, and grid operation. Furthermore, fairness measurements are incorporated into the planning process, allowing for the consideration of both horizontal transit equity and vertical transit equity based on different zone merging criteria within the county's existing census tracts. The results of this planning model offer valuable insights into achieving both economic efficiency and social justice in the design of on-route charging facilities for BEBs in South King County.
\end{abstract}


\maketitle
\begin{sloppypar}
\section{Introduction}\label{intro}
With a 27\% contribution to greenhouse gas emissions in 2020, the transportation system is the biggest economic sector that consumes fossil fuels \cite{us_epa_sources_2015}. To reduce the exhaust gas emissions of public transportation, the concept of electromobility, involving the adoption of electric vehicles (EVs) for transportation purposes, is rapidly being embraced by public transportation authorities. When it comes to bus systems, electromobility offers substantial advantages in terms of decreased operating and maintenance costs, increased energy efficiency, improved reliability, and reduced air and noise pollution \cite{topic2020virtual}. 

Over the course of recent decades, the global implementation of bus fleet electrification has emerged as a prominent and noteworthy trend. Notably, Shenzhen in China became the world's first city to fully electrify its public transit bus fleet in 2018, marking a historic achievement \cite{shenzhen_beb}. In Europe, the nations of the Netherlands and Luxembourg have made notable strides, with more than half of their registered city buses categorized as zero-emission vehicles \cite{netherlands_beb}. Similarly, King County in Washington, USA, has positioned itself as an early adopter of electric buses and is ambitiously transitioning towards a completely zero-emissions fleet by 2035 \cite{king_county_beb}. With remarkable advancements in battery technology, battery electric buses (BEBs) are becoming increasingly viable and appealing options for sustainable urban mobility, thus propelling cities worldwide toward a cleaner and more environmentally friendly future.

As bus agencies embrace this transition, they are driven by the dual objectives of ensuring economic efficiency and maintaining the service quality of their BEB fleets. Consequently, the optimization of charging infrastructure planning in this area becomes crucial, aiming to minimize investment and operation costs associated with the required charging facilities \cite{perumal2022electric}, as well as any additional costs that may arise during the electrification process.

A significant body of literature suggests that bus agencies often opt to construct charging facilities at designated base stations. In this approach, electric buses can only be charged after completing one or multiple full trips \cite{wang2017optimal, li2022multistage}, requiring them to deviate from their scheduled routes \cite{lin2019multistage} and travel deadheading distances for the purpose of charging \cite{rogge2018electric, lee2021optimal}. This off-route charging strategy is typically employed during overnight and layover periods when buses are not in service and have sufficient time for complete battery recharge \cite{mccabe2023optimal}. However, relying solely on this strategy may prove insufficient, especially in the case of King County, where the current on-base charging facilities can only meet 70\% of the bus assignments \cite{king_county_metro_king_2020}.

To bridge this energy gap, an alternative and promising direction to explore is the implementation of on-route charging stations. By strategically incorporating fast-charging facilities at on-street bus stops \cite{he2019fast, xylia2017locating}, BEBs can conveniently recharge during regular service operations. However, deploying on-route charging stations presents critical challenges that require attention. From an operational standpoint, limited research in BEB planning has explored the impact of the additional power load imposed by these on-route charging stations on the power grid \cite{elma2020flywheel}. This includes assessing power loss costs that may occur during electricity transmission. Furthermore, from a social perspective, the introduction of on-route charging stations must be approached with fairness in mind. Given that these stations can serve specific fixed routes \cite{liu2019economic}, it becomes imperative to ensure an equitable distribution of BEB routes across the regional transportation network. This ensures that diverse communities can access the associated benefits, such as cleaner air and enhanced environmental sustainability offered by BEBs.

Our proposed on-route fast-charging planning method effectively addresses the dual research gaps previously identified. Firstly, we prioritize the impacts on the local power grid in the placement of the on-route fast-charging infrastructure. To achieve this, we have developed a coupled power and transportation network specific to South King County. This integrated approach facilitates optimized planning, minimizing charging infrastructure investment and power system operational costs. Secondly, we recognize the limited attention given to equity in fleet electrification planning within the existing literature. To fill this gap, we have incorporated fairness measures into our planning approach to promote transit equity. Specifically, during the partial implementation of BEB routes in a particular region, our planning method promotes both horizontal and vertical transit equity by carefully selecting routes to be designated as BEB routes from the overall bus network.

The integration of the power and transportation networks, along with considerations of cost optimization and transit equity, positions our approach as an effective and comprehensive solution for the planning of on-route fast charging for BEBs. Furthermore, to emphasize the uniqueness of our method, we thoroughly examine existing research in fleet electrification planning, specifically focusing on the domains of power grid interaction and transit equity.

\subsection{Power Grid Interaction}\label{sec:pgi}
The successful implementation of fleet electrification necessitates a strong interconnection between transportation and power systems. To ensure efficient management of this interaction, an integrated approach that considers the coupled power and transportation network is crucial in charging infrastructure planning. While this approach has received limited attention in the context of electric bus on-route charging stations, some research has integrated the power grid and transportation network when planning EV charging stations. This integration can take two forms: coupling a transportation test case with a power system test case or coupling a real-world transportation network with a power system test case. The latter approach incorporates authentic data and conditions from a functioning transportation system, resulting in enhanced practicality.

In the first type of coupled system, the Sioux Falls network is widely used as a transportation test case. For example, in a study by \citet{he2013optimal}, a coupled network was created using the topology of the Sioux Falls network and a subset of the IEEE-118 bus system. The goal of this study was to allocate a specified number of charging stations for plug-in EVs. The potential locations of these charging stations were identified as common nodes in both the transportation and power grid systems. In another study by \citet{he2016sustainability}, the topology of the Sioux Falls network was retained for the transportation system, but the authors used a simplified version of the IEEE 34-bus system for the power grid. The authors matched the destination nodes in the transportation system with the corresponding buses in the power grid, but the intention behind this was unspecified. \citet{he2022comprehensive} built a coupled system using the Sioux Falls network and the IEEE 33-bus system; nevertheless, there was no direct relationship between the road distances and the power line lengths in this study.

In addition to the Sioux Falls network, other researchers have created their own transportation networks to build coupled systems for EV planning. For instance, \citet{he2013integrated} created a coupled system through the utilization of a nine-node road network and a subset of the IEEE 118-bus system. In this case, each link in the transportation network was connected to a particular bus in the power system, and the energy consumption of EVs on that link resulted in a power load on the grid. \citet{wang2013traffic} employed a 25-node traffic network and an 11 kV 33-node distribution system to construct their coupled system. The authors considered the geographical positioning of the nodes and established a direct relationship between the nodes in the transportation and power systems, where the traffic nodes 1-25 overlapped with the distribution system nodes 1-25. Furthermore, \citet{zhang2016pev} adopted a 25-node highway transportation network and designed a 14-node 110 kV high voltage distribution network to establish a relationship between the transportation link distance and the power line length within their coupled system.

Regarding the second type of coupled system, an exemplar is a work by \citet{lin2019multistage}, who employed a real-world transportation network from the city of Shenzhen and integrated it with the virtual power network established by \citet{zhang2016pev}. To retrieve distances within the transportation network, they utilized the API of Baidu Map. However, it is important to note that their studies did not account for the correlation between the actual road distance and the line length in the virtual power network.

Building upon the second type of coupled system discussed in the literature, we propose a comprehensive framework that integrates a real-world transportation network with a virtual power system. Our framework establishes a correlation between the actual bus route distance and the power line length in the coupled system, enhancing the practicality of the planning outcome. Unlike previous approaches, our framework is designed to be adaptable and suitable for various bus networks in different regions. By utilizing our generic coupled network framework, our objective is to address the existing research gap and provide a comprehensive solution for on-route BEB charging infrastructure planning.

\subsection{Transit Equity} 
Existing research has highlighted the presence of transit inequities among underserved communities, including people of color and low-income individuals, due to inadequate spatial coverage of transportation infrastructure \cite{mohl_stop_2004,venter_equity_2018}. Addressing and rectifying this long-standing spatial gap between low-income settlements and their access to transit services pose great challenges \cite{ermagun2023inequity}. However, fleet electrification, being a significant transit initiative, presents an opportunity to address these inequities right from the planning phase. This involves strategically locating charging infrastructure and designing efficient routes to serve historically underserved areas \cite{st_king_2022}. By adopting an equitable perspective, BEB planning \cite{king_county_metro_feasibility_2017} can serve as a means to mitigate the discriminatory impact on socially vulnerable populations caused by transit-related spatial mismatch. 

The concept of transit equity encompasses two dimensions: horizontal equity, promoting equal treatment for all individuals \cite{LitmanToddM2022ETEG}, and vertical equity, tailoring treatments to diverse needs or circumstances \cite{delbosc_spatial_2011}. Despite the importance of transit equity, there is a noticeable dearth of research that applies its principles to transportation-network-related planning \cite{camporeale2017quantifying}. \citet{fan2011bi} were the first to consider horizontal equity in solving the transportation network redesign problem by introducing a spatial equality constraint. Building upon this work, \citet{camporeale2017quantifying} combine both horizontal and vertical equity goals in a constraint of the transit network design problem, ensuring that the final configuration of the public transport service strikes the fairest compromise by considering both spatial distribution and social needs.

Furthermore, in the context of electric bus planning, there is even less research that incorporates transit equity. The work conducted by \citet{zhou2020bi} closely aligns with our research scope. They proposed a bi-objective model to support transit agencies in the optimal deployment of BEBs, taking into account capital investment and environmental equity. However, their primary focus lies in maximizing vertical equity in one of their objectives, which involves weighting disadvantaged populations based on air pollutant concentration. Notably, to the best of our knowledge, there have been no attempts to incorporate both horizontal and vertical equity into the planning problems of electric bus charging infrastructure. 

Given the limited research on the topic, it becomes necessary to draw upon metrics used in other domains to measure the fairness of the transit planning result. \citet{camporeale2017quantifying} employed the Gini coefficient, a widely used fairness metric in economics, to develop their equality constraint. Similarly, we have identified Jain's index, which is commonly used to measure fairness in resource allocation within telecommunication networks \cite{rezaeinia2022efficiency}, as a suitable metric to characterize the distribution of BEB routes across a bus network in a given region.

\subsection{Objective and Contribution}  
This paper presents a novel mixed-integer second-order cone programming (MISOCP) model that aims to optimize the placement of BEB on-route charging infrastructure. The objective is to minimize the planning and operation costs associated with the fleet electrification process in South King County, considering both the transportation and power systems. Additionally, we emphasize the importance of equity in the planning stage by incorporating fairness measurements, ensuring both horizontal and vertical equity. This research makes two primary contributions:

\begin{itemize}
    \item To address the potential challenge of BEB charging infrastructure on the power system effectively, we have implemented a coupled networks approach that integrates the local power grid and the bus networks into our planning model. Using South King County as a representative example, our proposed generic framework focuses on establishing a virtual power grid based on the under-planning bus network. By strategically deploying on-route charging stations at bus stops via solving the planning model, we establish coupling relationships between transportation nodes and power grid nodes, effectively integrating the two systems in the planning outcome. 
    
    \item To ensure equity in fleet electrification, we incorporate Jain's index as a fairness metric in our planning model for BEB charging infrastructure. In South King County, we aggregate census tracts based on both population and bus-commuter features, creating distinct subareas. By imposing a fairness constraint that ensures the desired level of Jain's index in these subareas, we promote equity in the planning results. The planning outcomes in the population-based subareas exhibit horizontal equity, ensuring an equal distribution of resources among all individuals. Conversely, the planning outcomes in the bus-commuter-based subareas demonstrate vertical equity, aiming for a fair allocation within the bus-commuter group.
    
\end{itemize}

The remainder of this paper is structured as follows. Section \ref{sec2} presents essential background information and prior knowledge on bus operation, coupled networks, and transit equity analysis in King County. This information is necessary for formulating the MISOCP model in Section \ref{sec3}. Section \ref{sec4} introduces a generic framework for establishing the coupled power and transportation network based on the given bus network, along with the corresponding algorithm. Case studies of the planning model, both with and without fairness measurement, are conducted in Section \ref{sec5}. Finally, the conclusion is drawn in Section \ref{sec6}.

\section{Problem Statement}\label{sec2}
The South Annex Base in King County is scheduled to open in 2025 and is expected to accommodate up to 250 BEBs \cite{king_county_metro_king_2020}. To ensure energy support for these vehicles, a combination of slower and faster on-base charging is planned to be adopted. Additionally, on-route fast charging is being considered as an augmented charging strategy for more frequent routes.

Among the various charging strategies for BEBs, there is a significant degree of flexibility in the deployment of on-route charging stations in different areas. This offers the opportunity to not only address the impacts of climate change but also to promote equity and social justice across the county. Thus, it is of utmost importance to determine the optimal location and capacity of on-route fast-charging stations for BEBs, such that the planning cost is minimized and fairness is maximized.

Furthermore, it is essential to consider the complex interplay between the transportation system and the power network in the planning process. The operation of fast-charging stations for BEBs will result in a significant load demand on the power grid, while at the same time providing stable energy services to BEBs on different routes. A comprehensive understanding of bus operation, the coupled network, and the transit equity status in King County is necessary to effectively implement the planning model.

\subsection{Bus Operation}
King County Metro conducted range tests on the 40-ft and 60-ft models of BEBs, validating their capacity to cover distances of up to 140 miles, which accounts for 70\% of the service needs \cite{king_county_metro_king_2020}. These buses are specifically designed for operation during morning and evening rush hours. By implementing on-route charging infrastructure, the BEBs can utilize smaller battery packs without compromising their operational effectiveness.

In the planning of on-route charging facilities, it is crucial to consider the existing on-base chargers at the Interim Base and explore cost-effective strategies for combining both charging strategies. While the current on-base chargers may not accommodate the extension of all new BEB routes, they do enable BEBs to be charged at varying initial state-of-charge (SOC) levels upon departure from their origin stations. Moreover, the implementation of on-route charging expands the range of the tested BEBs beyond the 140-mile capacity, enabling them to handle the remaining 30\% of vehicle assignments. This approach optimizes energy management for BEBs, ensuring they maintain sufficient charge to successfully complete their designated routes while minimizing the need for costly infrastructure upgrades.

\subsection{Network Representation}\label{sec:transrep}
In our forthcoming implementation of a coupled network framework for on-route charging station planning, we employ two symbolic systems to enhance the description of components within the transportation and power networks. This approach also simplifies the mathematical expression in our model formulation.

In regards to the transportation network, we denote the set of nodes as $N^T$ and the set of directed links as $L$. Within this network, a node $m\in N^T$ corresponds to a bus stop, and $\hat{N}^T$ is a subset of $N^T$ that represents the bus origin stations. For each bus route $\alpha\in \Omega_\alpha$, the route always starts from its origin station $o_\alpha$, where $o_\alpha \in \hat{N}^T$. The directed links and nodes that comprise bus route $\alpha$ are represented by $L_\alpha$ and $N_\alpha^T$, respectively.

Concerning the power grid, we denote the set of nodes as $N^G$ and the set of branches as $\mathcal{E}$. In this context, a node in the power grid is represented by $i\in N^G$, while a power line is denoted as $(i,j)\in \mathcal{E}$.

\subsection{Equity Analysis}
\begin{figure}[!h]
\centering
\includegraphics[width=0.9\linewidth]{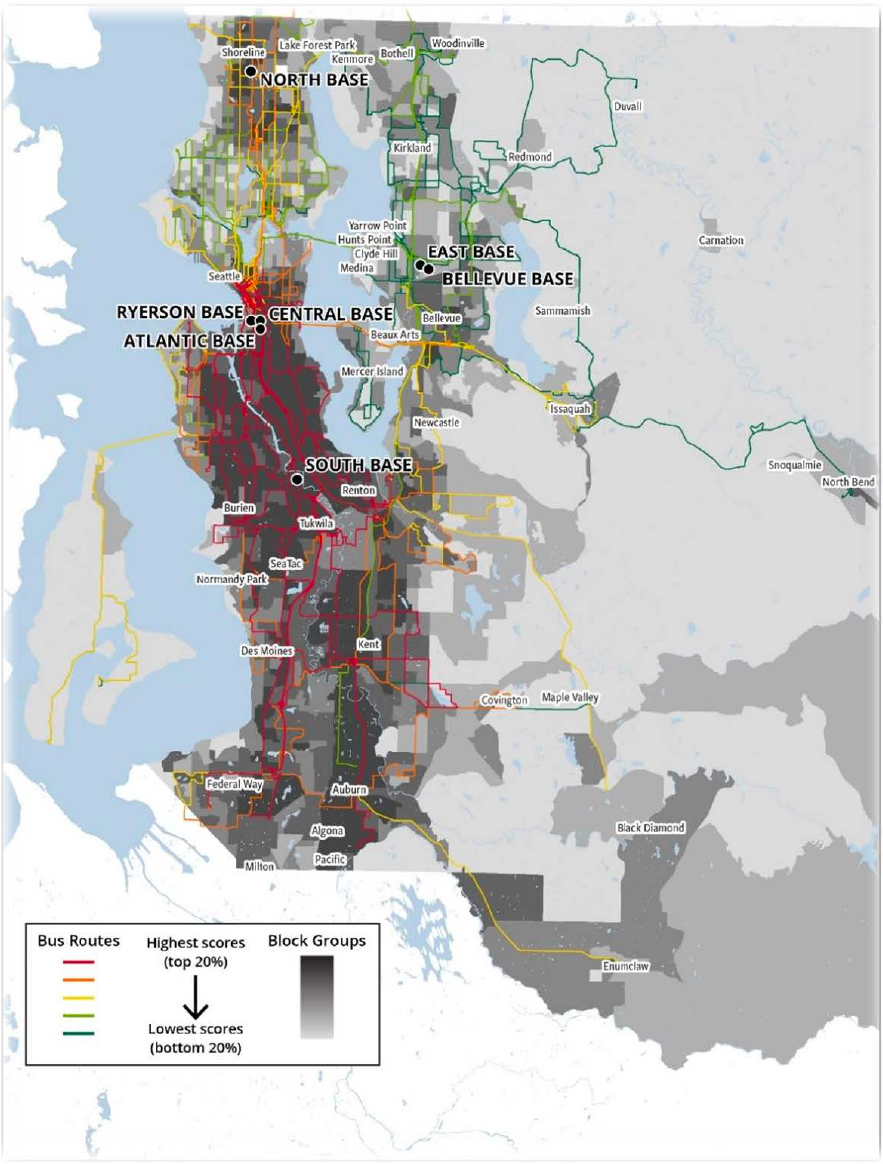}
\caption{Map of air pollution vulnerable areas and priority quintiles for zero-emission bus service in King County\protect\footnotemark[1].}
\label{fig1}
\end{figure}

Neighborhoods in King County that are burdened with elevated air pollution levels tend to be home to low-income households and marginalized racial and ethnic groups \cite{king_county_metro_king_2020}. This disparity is demonstrated in Figure \ref{fig1}, where the southern base areas, encompassing Renton, Burien, Tukwila, SeaTac, and Kent, are prominently affected. These communities have long endured inadequate transportation services, resulting in heightened exposure to transportation-related noise and air pollution. Notably, the South Base exhibits a higher number of daily service miles compared to other bases, indicating a greater extent of service inadequacy. Moreover, approximately 31\% of the census blocks along South Base routes are categorized as highly vulnerable to the adverse impacts of air pollution \cite{king_county_metro_feasibility_2017}.

\footnotetext[1]{Data Source: King County Metro’s report on Feasibility of Achieving a Carbon-Neutral or Zero-Emissions Fleet}

By prioritizing the implementation of BEB plans in South King County, we can maximize transit equity countywide. The introduction of zero-emission bus routes significantly improves air quality and public health, particularly benefiting minority communities. Moreover, BEBs offer a more comfortable commuting experience with smoother and quieter acceleration and deceleration \cite{quarles2020costs}. This improved ride quality is valuable for daily bus commuters, especially low-income individuals who heavily depend on buses. By enhancing their overall experience, we have the potential to increase ridership and improve public transportation accessibility, thus advancing social equity objectives in fleet electrification planning.

\section{Model Formulation}\label{sec3}
In this section, we present the formulation of a mathematical model designed for BEB on-route charging station planning. We delve into the details of incorporating Jain's index as a fairness measure within the planning framework. The model optimization includes determining the optimal placement of charging stations, the number of charging piles at each station, the interconnection between bus stations and power grid nodes, and the current flow through power lines that connect the bus stations to the power grid.

\subsection{Objective Function}
The total planning cost for the on-route charging infrastructure of BEBs is determined by considering the costs associated with both the transportation network for constructing the facilities and the power grid for integrating the new charging stations. This cost is represented by \eqref{obj}, which consists of four components that are summed together. The first three components pertain to the investment cost for the charging stations, charging piles, and the power lines connecting the charging stations to the power grid. The final component represents the operational cost, which accounts for the energy loss in the power grid integrated with on-route charging stations.

\begin{equation}
\begin{aligned}
\min &\sum_{m\in N^T}(f_{s,m}\cdot X_m + f_{c,m}\cdot \beta_m) + \sum_{m\in N^T}\sum_{i\in N^G} c_{i,m}\cdot \mathit{\Psi}_{i,m} \\
&+ \sum_{\mathcal{E}:(i.j)}T\cdot c_e\cdot \ell_{ij}\cdot r_{ij},  \label{obj}
\end{aligned}
\end{equation}
where $f_{s,m}$ and $f_{c,m}$ stand for the unit cost of charging stations and charging piles at bus station $m$, respectively. The binary decision variable $X_m$ represents the construction of a charging station at bus station $m$, with $X_m=1$ signifying its presence. The integer decision variable $\beta_m$ represents the number of charging piles installed at bus station $m$. The cost of constructing the power line that connects bus station $m$ to the power grid node $i$ is represented by $c_{i,m}$, which is determined by the geographical distance between the two nodes. The binary decision variable $\mathit{\Psi}_{i,m}$ indicates whether the power line has been established, where $\mathit{\Psi}_{i,m}=1$ signifies that bus station $m$ has been successfully integrated into power grid node $i$. The power loss time in the planning period is represented by $T$, while the electricity price is $c_e$. $\ell_{ij}$ denotes the square of the magnitude of the complex current from node $i$ to $j$ after building charging stations, while $r_{ij}$ represents the resistance of power line $(i,j)$.

\subsection{Constraints}\label{sec:cons}
The introduced constraints in the planning model cover three essential processes: ensuring that BEB batteries have sufficient charge to complete their routes, managing the energy transfer from the power grid to the BEB batteries at the charging stations, and assessing the influence of integrated charging stations on the local power grid's power flow. These constraints also establish the coupling relationship between the bus stops in the transportation network and their corresponding nodes in the power grid.

In order for a bus to be charged at bus station $m$, it is mandatory for a charging station to be constructed at that location:
\begin{equation}
    y_{\alpha,m}\le X_m, \ \ \forall \alpha\in \Omega_\alpha,\forall m\in N^T, \label{cons:oper}
\end{equation}
where $y_{\alpha,m}$ is a binary decision variable, and $y_{\alpha,m} = 1$ denotes the bus on route $\alpha$ charges at bus station $m$. In addition, it is necessary to construct the charging piles at an established charging station, which can be formalized as follows using a Big-M method:
\begin{equation}
    0\le \beta_m \le M \cdot X_m, \ \ \forall m\in N^T, 
\end{equation}
where $M$ can be considered as the total number of available charging piles to be invested during the planning period.

To avoid any queuing during the limited on-route charging time slots, a practical approach is to assign dedicated charging piles for each bus route at shared stations. This strategy ensures smooth charging operations and minimizes potential disruptions or delays caused by congested charging stations. Therefore, it is necessary to ensure that the number of installed charging piles is no less than the number of bus routes assigned to charge at the station:
\begin{equation}
    \sum_{\alpha\in \Omega_\alpha} y_{\alpha,m} \le \beta_m, \ \ \forall m\in N^T.
\end{equation}

In order to establish a functional coupling between the transportation and power network, it is imperative to connect bus stations that are selected for installing charging infrastructure to a power grid node:
\begin{equation}
    \sum_{i\in N_G} \mathit{\Psi}_{i,m} = X_m, \ \ \forall m\in N^T. \label{cons:connect}
\end{equation}

Given the requirement for all BEBs to complete their round trips successfully, we consider the initial SOC of their batteries, deviating from previous studies \cite{he2019fast, liu2018planning} that assume fully-charged batteries at the start. By exploring various levels of initial SOC as BEBs depart from their origin stations, we can determine the corresponding optimal scale of on-route charging facilities. As a result, we can effectively manage the investment in on-route charging facilities and make efficient use of the existing on-base charging stations at the Interim Base. The initial energy of the BEBs at the time of departure from the origin station $o_\alpha$ can be quantified as $\theta_0 \cdot u_\alpha^{bat}$, where $\theta_0$ represents the initial SOC of the batteries, and $u_\alpha^{bat}$ denotes the specific battery capacity of bus route $\alpha$.

During BEB operation, it is necessary to maintain the battery's SOC within a specific safe range:
\begin{align}
    e_{\alpha,m} &\ge \theta^l \cdot u_\alpha^{bat},\ \ \forall \alpha\in \Omega_\alpha,\forall m\in N^T, \label{cons:e_lower}\\
    e_{\alpha,m} + s_{\alpha,m} &\le \theta^u \cdot u_\alpha^{bat},\ \ \forall \alpha\in \Omega_\alpha,\forall m\in N^T, 
\end{align}
where $\theta^l$ and $\theta^u$ are the lower and upper bound of the battery capacity of BEBs. $e_{\alpha,m}$ and $s_{\alpha,m}$ represent the energy level and the battery's energy supply in BEBs for route $\alpha$ at the station $m$.

At each bus station, the energy conservation constraint for BEB batteries accounts for the energy consumption during travel between stations $m$ and $n$:
\begin{equation}
    e_{\alpha,n} = e_{\alpha,m} + s_{\alpha,m} - e^0_\alpha\cdot d_{mn},\ \ \forall \alpha\in \Omega_\alpha,\forall (m,n)\in L_\alpha, \label{cons:energy_cons}
\end{equation}
where $e^0_\alpha$ denotes the average energy consumption of BEBs per unit distance for route $\alpha$, which depends on the specific BEB model used. The driving distance between stations $m$ and $n$ is represented by $d_{mn}$. It is worth noting that we consider round-trip routes for each BEB, and the bus must satisfy the energy conservation constraint during the completion of its route in both directions.

All BEBs are required to adhere to the predefined operation schedule and cannot spend excessive time at a charging station. Therefore, the charging energy must not exceed the maximum available energy supply:
\begin{equation}
    0\le s_{\alpha,m}\le P^e_\alpha \cdot \tau_{\alpha,m}\cdot y_{\alpha,m}, \ \ \forall \alpha\in \Omega_\alpha,\forall m\in N^T, \label{cons:max_supply}
\end{equation}
where $P^e_\alpha$ denotes the nominal power of the charging pile for bus route $\alpha$, and $\tau_{\alpha,m}$ represents the maximum dwelling time for bus route $\alpha$ at station $m$.

To determine the actual power loss in the power grid after incorporating on-route charging stations, we can utilize a branch flow model as described in \citet{farivar2013branch}:
\begin{flalign}
    s_j &= \sum_{k:j\to k}S_{jk} - \sum_{i:i\to j}(S_{ij} - z_{ij}\ell_{ij}),  \ \ \forall (i,j)\in \mathcal{E}, &&\label{cons:opf_e1}\\
    v_j &= v_i - 2(r_{ij}P_{ij}+x_{ij}Q_{ij})+(r_{ij}^2+x_{ij}^2)\cdot\ell_{ij}, \ \ \forall (i,j)\in \mathcal{E}, &&\label{cons:opf_e2} \\
    \ell_{ij} &= \frac{P_{ij}^2+Q_{ij}^2}{v_i}, \ \ \forall (i,j)\in \mathcal{E}, &&\label{cons:opf_e3}      
\end{flalign}
where $s_j$ represents the power injection at power grid node $j$. $S_{ij}$ denotes the sending-end power flow from node $i$ to $j$, given by $S_{ij} = P_{ij} + \mathbf{i}Q_{ij}$. $z_{ij}$ is the impedance of line $(i,j)$, represented as $z_{ij} = r_{ij} + \mathbf{i}x_{ij}$. $\ell_{ij}$ represents the square of the magnitude of the complex current from node $i$ to $j$, while $v_j$ represents the square of the magnitude of the complex voltage at node $j$. The resistance and reactance of line $(i,j)$ are represented by $r_{ij}$ and $x_{ij}$, respectively. Furthermore, the real power flow from node $i$ to node $j$ is denoted as $P_{ij}$, and $Q_{ij}$ signifies the reactive power flow between these nodes.

The power injection at a power grid node in \eqref{cons:opf_e1} consists of two components: the charging power from integrated on-route charging stations, if applicable, and the original load demand:
\begin{equation}
    s_i = -\sum_{m\in N^T} \sum_{\alpha\in\Omega_\alpha} P^e_\alpha\cdot y_{\alpha,m} \cdot \mathit{\Psi}_{i,m} - s^{load}_i, \ \ \forall i\in N^G, \label{cons:powerinj}
\end{equation}
where $s^{load}_i$ is the original load demand of power node $i$.

For the reliable and safe operation of the power grid after integrating on-route charging stations, it is crucial to maintain both the voltage and current within a specific range:
\begin{align}
    \underline{v_i}\le v_i\le \overline{v_i}, \ \ \forall i\in N^G, \label{cons:nodevol}\\
    0\le \ell_{ij}\le \overline{\ell_{ij}},\ \ \forall (i,j)\in \mathcal{E},
\end{align}
where $\underline{v_i}$ and $\overline{v_i}$ denote the lower and upper bound of the square of the node voltage, respectively. $\overline{\ell_{ij}}$ represents the maximum square of the current in line $(i,j)$.

Finally, we ensure that all binary and integer decision variables used in the planning model satisfy the following conditions:
\begin{align}
    X_m &\in \{0,1\},\ \ \forall m\in N^T, \\
    \beta_m &\in \mathbb{Z},\ \ \forall m\in N^T, \\
    y_{\alpha,m} &\in \{0,1\},\ \ \forall \alpha \in \Omega_\alpha,\forall m\in N^T, \\
    \mathit{\Psi}_{i,m} &\in \{0,1\},\ \ \forall i \in N^G,\forall m\in N^T.
\end{align}

\subsection{Model Relaxations}
In order to make the planning model compatible with commercial solvers like Gurobi and CPLEX, certain nonlinear constraints in Section \ref{sec:cons} need to be relaxed. The first constraint to be handled with is \eqref{cons:opf_e3} due to its quadratic term. To address this, we adopt the approach proposed by \citet{farivar2013branch} and reformulate it as the following second-order cone constraint:
\begin{equation}
    \begin{Vmatrix}
        2P_{ij}\\
        2Q_{ij}\\
        \ell_{ij}-v_i
    \end{Vmatrix}_2 \le \ell_{ij}+v_i, \ \ \forall (i,j)\in \mathcal{E}.
\end{equation}

Another non-linearity lies in \eqref{cons:powerinj}, which involves the product of two binary decision variables $y_{\alpha,m}$ and $\mathit{\Psi}_{i,m}$. We introduce an auxiliary variable $Y_{\alpha,m,i}$ to replace the product. Constraint \eqref{cons:powerinj} is then reformulated as follows:
\begin{equation}
    s_i = -\sum_{m\in N^T} \sum_{\alpha\in\Omega_\alpha} P^e_\alpha\cdot Y_{\alpha,m,i} - s^{load}_i, \ \ \forall i\in N^G. \label{cons:powerinj_v2}
\end{equation}
To ensure the consistency between the auxiliary variable $Y_{\alpha,m,i}$ and the product of $y_{\alpha,m}$ and $\mathit{\Psi}_{i,m}$, we introduce additional constraints:
\begin{flalign}
    Y_{\alpha,m,i} &\le y_{\alpha,m}, \ \ \forall \alpha \in \Omega_\alpha,\forall m\in N^T,\forall i\in N^G,&&\\
    Y_{\alpha,m,i} &\le \mathit{\Psi}_{i,m}, \ \ \forall \alpha \in \Omega_\alpha,\forall m\in N^T,\forall i\in N^G,&&\\
    Y_{\alpha,m,i} &\ge y_{\alpha,m}+\mathit{\Psi}_{i,m}-1,\ \ \forall \alpha \in \Omega_\alpha,\forall m\in N^T,\forall i\in N^G,&&\\
    Y_{\alpha,m,i} &\in \{0,1\}, \ \ \forall \alpha \in \Omega_\alpha,\forall m\in N^T,\forall i\in N^G.&& \label{cons:auxi}
\end{flalign}

Consequently, the entire planning model is reformulated as a MISOCP problem:
\begin{equation}
    \begin{aligned}
        \min &\sum_{m\in N^T}(f_{s,m}\cdot X_m + f_{c,m}\cdot \beta_m) + \sum_{m\in N^T}\sum_{i\in N^G} c_{i,m}\cdot \mathit{\Psi}_{i,m} \\
        &+ \sum_{\mathcal{E}:(i.j)}T\cdot c_e\cdot \ell_{ij}\cdot r_{ij} \\
        \text{s.t.} \  & \eqref{cons:oper}-\eqref{cons:opf_e2}, \eqref{cons:nodevol}-\eqref{cons:auxi}. \label{planning_wo_fair}
    \end{aligned}
\end{equation}

\subsection{Fairness Measures}
In our model, we utilize Jain's index \cite{jain1984quantitative} as a measure of fairness in the planning of on-route charging stations for BEBs. Jain's index possesses several desirable properties, including population size independence, scale and metric independence, boundedness, and continuity. If we divide the planning area in South King County into $H$ areas and assign an allocation of $w_h$ to the $h$th area, then the expression for Jain's index can be given as follows:
\begin{equation}
    f(w) = \frac{(\sum_{h=1}^H w_h)^2}{H\sum_{h=1}^H w_h^2}.\label{eq:jain}
\end{equation}

To determine the fairness index $w_h$ in our planning model, we must consider the impact of a zero-emission fleet on the residents of King County. Figure \ref{fig1} emphasizes the southern regions, marked by the red bus routes and dark-shaded areas, which are disproportionately affected by air pollution and inadequate transit service. The deployment of zero-emission buses in these areas would provide significant equity benefits. Hence, the fairness index $w_h$ should reflect the reduction in air pollution and the improvement in traditional bus services resulting from our BEB planning efforts.

\begin{figure}[!h]
\centering
\includegraphics[width=0.7\linewidth]{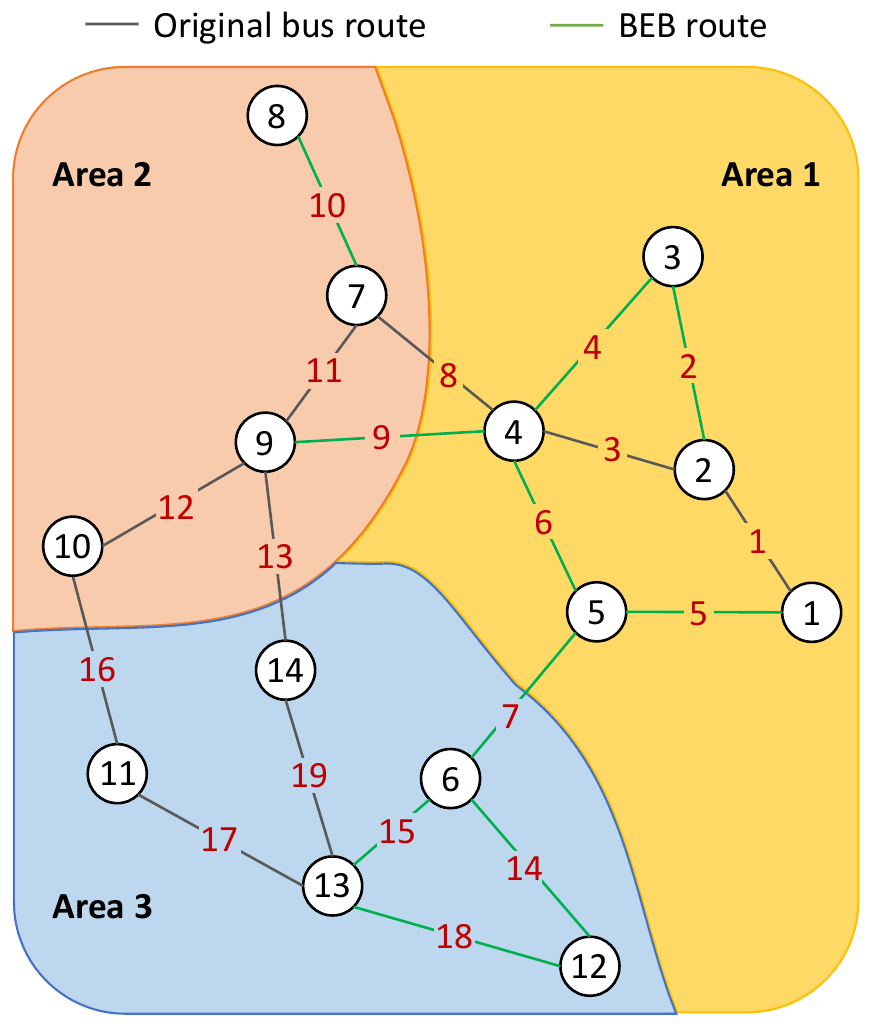}
\caption{Illustration of the definition of fairness index $w_h$.}
\label{fig2}
\end{figure}  
A viable approach to establishing such a fairness index would be to consider the proportion of BEB routes within a specific area relative to all bus routes. In Figure \ref{fig2}, we present a simplified transportation network comprising 14 nodes and 19 directed links. These links are divided into three distinct areas, with the assumption that links crossing multiple areas are evenly distributed. We define the fairness index $w_h$ as the ratio of the total length of BEB routes to the total length of all bus routes within each area. The ratios for BEB routes in areas 1, 2, and 3 are denoted as $w_1$, $w_2$, and $w_3$ respectively, and can be computed using the following equations:
\begin{equation}
\begin{split}
 w_1 &= \frac{d_2+d_4+d_5+d_6+0.5d_7+0.5d_9}{d_1+d_2+d_3+d_4+d_5+d_6+0.5d_7+0.5d_8+0.5d_9}, \\
 w_2 &= \frac{0.5d_9+d_{10}}{0.5d_8+0.5d_9+d_{10}+d_{11}+d_{12}+0.5d_{13}+0.5d_{16}}, \\
 w_3 &= \frac{0.5d_7+d_{14}+d_{15}+d_{18}}{0.5d_7+0.5d_{13}+d_{14}+d_{15}+0.5d_{16}+d_{17}+d_{18}+d_{19}},
\end{split}
\end{equation} 
where $d_l$ $(l=1,\cdots, 19)$ denotes the driving distance along each link $l$ of the bus route, as shown in the figure. This approach ensures that the fairness index $w_h$ within each area remains independent of other factors, including population density, the number of routes, or the size of the area.

To explicitly express the fairness index as a percentage of BEB routes in our planning model, we introduce a new binary decision variable $I_\alpha$. Here, $I_\alpha = 1$ indicates that bus route $\alpha$ is selected as a BEB route. Once all directed links $(m,n)\in L$ are assigned to the $H$ areas, we define $L_A^h$ as the set of all links in the $h$th area. Consequently, we have the following relationship:
\begin{equation}
    w_h = \frac{\sum_{\alpha\in \Omega_\alpha}\sum_{L:(m,n)\in L_A^h \cap L_\alpha}d_{mn}\cdot I_\alpha}{\sum_{L:(m,n)\in L_A^h}d_{mn}}, \ \ \forall h=1,\dots, H. \label{cons:wh_eq}
\end{equation}

Considering the bounded nature of Jain's index as defined in \eqref{eq:jain}, we can observe that $f(w)$ satisfies the inequality $\frac{1}{H} \leq f(w) \leq 1$. As the value of $f(w)$ increases, the fairness level also increases, reaching maximum fairness when $f(w)=1$ (100\% fair). To ensure a desired fairness level, we introduce a constraint as follows:
\begin{equation}
    f(w) = \frac{(\sum_{h=1}^H w_h)^2}{H\sum_{h=1}^H w_h^2} \ge \eta, \label{cons:fair}
\end{equation}
where $\eta$ denotes a predetermined fairness level of the planning result, constrained to be between $\frac{1}{H}$ and $1$.

Note that the quadratic term in \eqref{cons:fair} results in non-linearity, which can be reformulated in the following manner:
\begin{equation}
    \sum_{h=1}^H w_h^2\le \frac{1}{H\cdot \eta}(\sum_{h=1}^H w_h)^2. \label{cons:fair_ieq}
\end{equation}
This inequality constraint can be further rewritten as a second-order cone constraint:
\begin{equation}
    \begin{Vmatrix}
        w_1\\
        \vdots \\
        w_H
    \end{Vmatrix}_2 \le \sum_{h=1}^H \sqrt{\frac{1}{H\cdot \eta}} w_h. \label{cons:fair_socp}
\end{equation}

The introduction of $I_\alpha$ necessitates the reformulation of certain constraints in Section \ref{sec:cons}. First, it enables us to quantify the initial energy $e_{\alpha,o_\alpha}$ saved in all BEB batteries as below:
\begin{equation}
    e_{\alpha,o_\alpha} = \theta_0 \cdot u_\alpha^{bat}\cdot I_\alpha, \ \ \forall \alpha\in \Omega_\alpha. \label{cons:news0}
\end{equation}
In a similar fashion, we can redefine constraints \eqref{cons:e_lower}-\eqref{cons:energy_cons} as follows:
\begin{flalign}
    e_{\alpha,m} &\ge \theta^l \cdot u_\alpha^{bat} \cdot I_\alpha,\ \ \forall \alpha\in \Omega_\alpha,\forall m\in N^T,&&\label{cons:newe_lower}\\
    e_{\alpha,m} &+ s_{\alpha,m} \le \theta^u \cdot u_\alpha^{bat} \cdot I_\alpha,\ \ \forall \alpha\in \Omega_\alpha,\forall m\in N^T, &&\\
    e_{\alpha,n} &= e_{\alpha,m} + s_{\alpha,m} - e^0\cdot d_{mn} \cdot I_\alpha,\ \ \forall \alpha\in \Omega_\alpha,\forall (m,n)\in L_\alpha. &&\label{cons:newenergy_cons}
\end{flalign}
Regarding BEB routes, when $I_\alpha = 1$, constraints \eqref{cons:newe_lower}-\eqref{cons:newenergy_cons} are equivalent to \eqref{cons:e_lower}-\eqref{cons:energy_cons}. On the other hand, for non-BEB routes where $I_\alpha = 0$, we have $e_{\alpha,o_\alpha} = e_{\alpha,m} = s_{\alpha,m}=0$.

Furthermore, additional constraints need to be incorporated to account for the new decision variable $I_\alpha$, which ensures that buses can only charge if their routes are designated for BEBs:
\begin{equation}
    y_{\alpha,m}\le I_\alpha, \ \ \forall \alpha\in \Omega_\alpha,\forall m\in N^T.
\end{equation}
And if a bus on route $\alpha$ never undergoes charging, it indicates that the route is not intended for BEBs.
\begin{equation}
    I_\alpha \le \sum_{m\in N^T}y_{\alpha,m}, \ \ \forall \alpha\in \Omega_\alpha.
\end{equation}

Considering the budget limitations associated with constructing on-route charging facilities, we impose an upper limit on the number of BEB routes:
\begin{equation}
    \sum_{\alpha\in \Omega_\alpha}I_\alpha \le I_{\text{max}},
\end{equation}
where $I_{\text{max}}$ represents the maximal number of BEB routes to be invested in the planning period. And we introduce the following binary constraint:
\begin{equation}
    I_\alpha \in \{0,1\}, \ \ \forall \alpha\in \Omega_\alpha. \label{cons:binaryI}
\end{equation}

As a result, the model formulation that takes fairness measurement into consideration remains a MISOCP problem:
\begin{equation}
    \begin{aligned}
        \min &\sum_{m\in N^T}(f_{s,m}\cdot X_m + f_{c,m}\cdot \beta_m) + \sum_{m\in N^T}\sum_{i\in N^G} c_{i,m}\cdot \mathit{\Psi}_{i,m} \\
        &+ \sum_{\mathcal{E}:(i.j)}T\cdot c_e\cdot \ell_{ij}\cdot r_{ij} \\ 
        \text{s.t.} \  & \eqref{cons:oper}-\eqref{cons:connect}, \eqref{cons:max_supply}-\eqref{cons:opf_e2}, \eqref{cons:nodevol}-\eqref{cons:auxi}, \eqref{cons:wh_eq}, \eqref{cons:fair_socp}-\eqref{cons:binaryI}. \label{planning_with_fair}
    \end{aligned}
\end{equation}

\section{Coupled Network Framework}\label{sec4}

As outlined in Section \ref{sec:pgi}, the electrification of bus fleets involves two types of coupled power and transportation systems. Our study adopts the second type, whereby a real-world transportation network in South King County is integrated with a virtual distribution power network. Initially, we acquired the transportation map of the potential electric bus routes in South King County, which served as the basis for designing a virtual power network. Subsequently, our study aimed to establish a practical connection between these two networks.

\begin{figure}[!h]
\centering
\includegraphics[width=\linewidth]{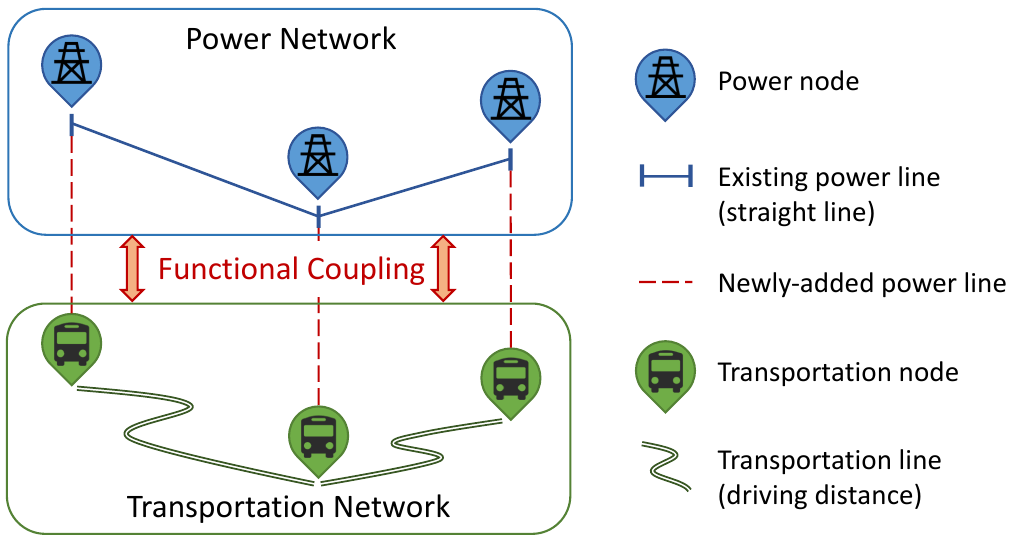}
\caption{Functional interconnection of nodes in the power and transportation networks.}
\label{fig3}
\end{figure}

It is important to highlight that our coupled network framework differs from existing research studies \cite{wang2013traffic, zhang2016pev} in that we did not assume an equivalence between the transportation links and the power line lengths in the coupled system. Instead, we determine the transportation link based on the driving distance between two transportation nodes, while the length of the power line is determined by the straight-line distance between power grid nodes. As illustrated in Figure \ref{fig3}, our approach involves constructing on-route charging stations at existing bus stations that represent transportation nodes. And the newly-added power lines invested in \eqref{obj}, depicted as red dashed lines, facilitate the functional coupling between the transportation network and the power network. These power lines efficiently transmit electrical energy, ensuring a stable and reliable bus service across the entire transportation network.
This approach enhances the practical significance of the links in both networks, enabling more accurate and efficient calculations in the planning problem.

\subsection{Transportation Network Design}\label{sec:transnet}
We construct a transportation network using the electric bus routes in South King County, with the on-route bus stations serving as transportation nodes and the route segments serving as transportation links within the network. To identify the potential routes, we refer to Appendix C of the King County Transit report \cite{king_county_metro_king_2020} and exclude the non-operational bus routes. The remaining routes, including 22, 101, 102, 111, 150, 153, 156, 168, 177, 181, 182, 183, 187, 190, and 193, are used to establish the transportation network. This network is derived from the general transit feed specification (GTFS) data \cite{kingcountygov_gtfs}.

When abstracting the transportation network from the intricate bus route map, it is essential to consider that using all bus stations along the identified bus routes as nodes in the transportation network may not be feasible due to computational complexity. Therefore, we have developed a set of rules for selecting the nodes from the available bus stations. These rules are designed to take into account practical considerations and are as follows:
\begin{enumerate}[label={\arabic*)}]
    \item When selecting nodes, priority is assigned to common bus stops that are connected to multiple bus routes, as they have a greater impact on the transportation network.
    \item For each bus route, the origin station and terminal station are identified as nodes in the transportation network.
    \item A distance threshold is set starting from the origin station. During the node selection process, we assess the distance between the current bus station and the previously selected node. If this distance exceeds the threshold, we incorporate the station preceding the current bus station as a new node within the transportation network.
    \item The service loop of a bus is also taken into account. Bus stations located on opposite sides of the street are considered separate nodes if both are selected.
\end{enumerate}

\begin{algorithm}
\DontPrintSemicolon
\caption{Transportation Node Selection}\label{alg:transnode}
\KwData{$\mathcal{S}_{\text{all}}$, $\mathcal{S}_\alpha$, $d_{0,s}$, and $d_\theta^T$}
\KwResult{selected transportation nodes $N^T$}
$N^T \gets \emptyset$\;
$\mathcal{S}_{\text{count}} \gets$ an array of zeros with length $|\mathcal{S}_{\text{all}}|$\;
 \For{$\alpha \in \Omega_\alpha$}{
  \tcc{selection rule 2)}
  $N^T \gets N^T \cup \{\mathcal{S}_\alpha[0]\}$\;
  $N^T \gets N^T \cup \{\mathcal{S}_\alpha[-1]\}$\;
  $\Delta d \gets 0$\;
    \For{$s \in \mathcal{S}_\alpha$}{
    $\mathcal{S}_{\text{count}}[s] \gets \mathcal{S}_{\text{count}}[s] + 1$\; 
    \tcc{selection rule 3) \& 4)}
    \If{$d_{0,s} > \Delta d + d_\theta^T$}{
      \eIf{$d_{0,s} - d_{0,s^-} > d_\theta^T$}{
       $N^T \gets N^T \cup \{s^-, s\}$\;
       $\Delta d \gets d_{0,s}$\;
       }{
       $N^T \gets N^T \cup \{s^-\}$\;
       $\Delta d \gets d_{0,s^-}$\;
      }
    }   
    }
 }
 \For{$s \in \mathcal{S}_{\text{\textnormal{all}}}$}{
    \tcc{selection rule 1)}
    \If{$\mathcal{S}_{\text{\textnormal{count}}}[s] > n_{\text{\textnormal{count}}}$}{
    $N^T \gets N^T \cup \{s\}$\;
    }
 }
 \KwRet{$N^T$}
\end{algorithm}

Algorithm \ref{alg:transnode} outlines the methodology for selecting transportation nodes from all bus stations to identify potential locations for building on-route charging stations. The set of all bus stations from the 15 identified bus routes is denoted as $\mathcal{S}_{\text{all}}$, while $\mathcal{S}_\alpha$ represents on-route bus stations along a specific route $\alpha$. We represent the initial and final bus stations of route $\alpha$ as $\mathcal{S}_\alpha[0]$ and $\mathcal{S}_\alpha[-1]$, respectively. Notably, $\mathcal{S}_\alpha[0]$ corresponds to the origin station $o_\alpha$ defined in Section \ref{sec:transrep}. The current and preceding bus station IDs are designated as $s$ and $s^-$ respectively. Additionally, the number of occurrences of each bus station for all identified bus routes is recorded in $\mathcal{S}_{\text{count}}$. $d_{0,s}$ indicates the cumulative driving distance of bus station $s$ from its origin station, and $\Delta d$ is the cumulative driving distance of the previously selected transportation node from its origin station. The distance threshold set in selection rule 3) is represented by $d_\theta^T$, while $n_{\text{count}}$ denotes the minimum number of bus routes that will be served by the common bus stop as defined in selection rule 1).

The algorithm begins by initializing the selected transportation node-set and the occurrence of each station in lines 1-2. Then, lines 4-5 iterate over all routes, adding the origin and terminal bus stations to the $N^T$ set as requested by selection rule 2). For each bus station $s$ along the route, its occurrence count is incremented by 1 in line 8, and lines 9-10 compare the distance between the current bus station and the previously selected node to the distance threshold $d_\theta^T$. If the distance exceeds the threshold, the algorithm selects the station before the current bus station as a new node in the transportation network, as requested by selection rule 3).

Once the occurrences of all bus stations have been counted, lines 21-25 check if the number of bus routes that serve the station $s$ exceeds $n_{\text{count}}$. If so, the station is added to the $N^T$ set in accordance with selection rule 1). The algorithm also considers selection rule 4) by using the parameter $\mathcal{S}_\alpha$, which collects the bus stations along a round trip journey of bus route $\alpha$ in the order of their actual driving cycle.
\begin{figure}[!h]
\centering
\includegraphics[width=\linewidth]{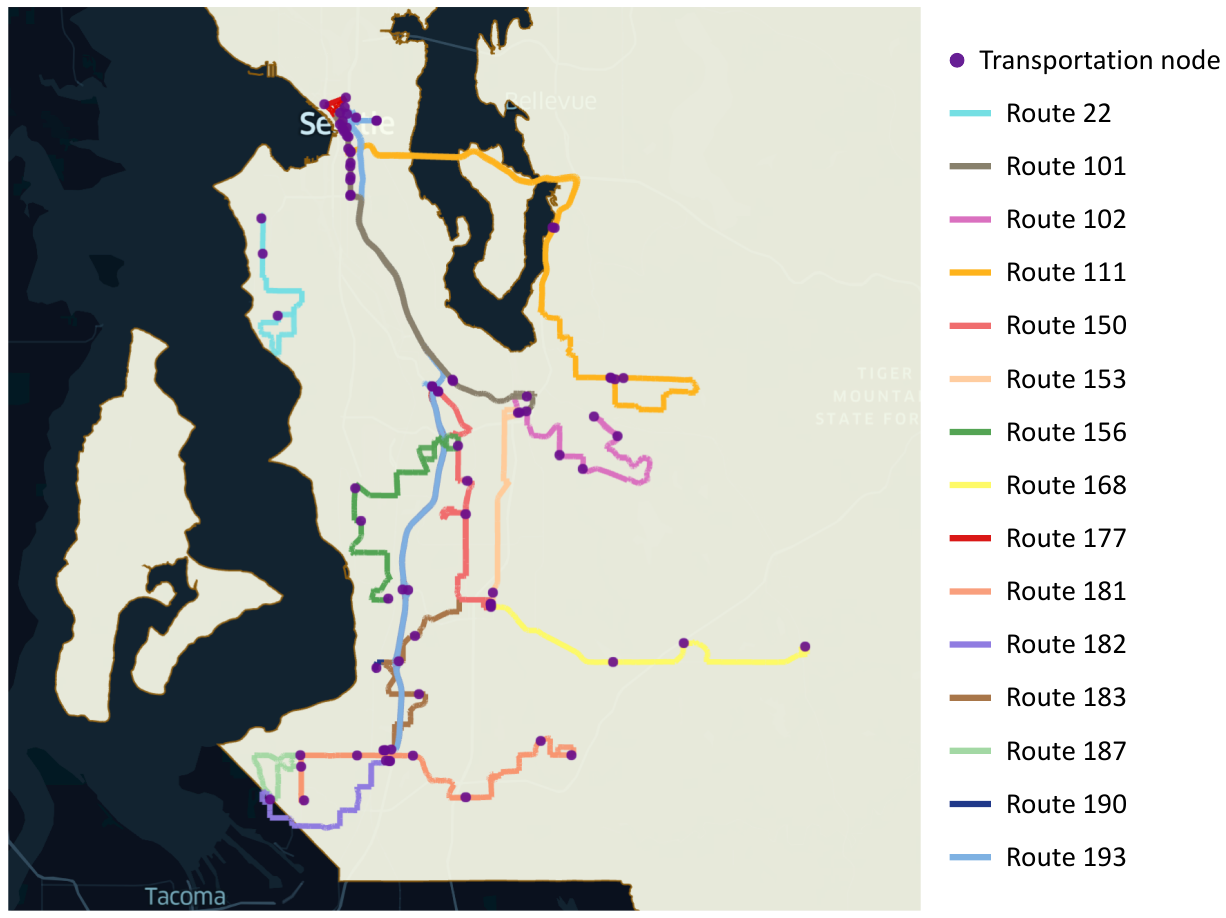}
\caption{Transportation network of 15 BEB routes in South King County.}
\label{fig4}
\end{figure}  

By applying the selection criteria and algorithm, we have effectively generated a transportation network, as depicted in Figure \ref{fig4}. This network comprises the 15 currently available BEB routes, which are indicated by different colors, and the selected bus stations are represented by purple points. In this study, we defined a common bus stop as a station that serves more than three bus routes $(n_{\text{count}} = 3)$. We set the driving distance threshold $d_\theta^T$ to 40,000 ft and identified a total of 84 bus stations as network nodes, each with the potential to accommodate on-route charging station installations.

\subsection{Coupled Virtual Power Network Design}
Using the transportation network depicted in Figure \ref{fig4} as a foundation, we construct a virtual power network that aligns geographically with the selected bus stations. However, the distance between two adjacent bus stations is usually much shorter than the distance between two power grid nodes in reality. Therefore, we need to further refine the selection of power grid nodes from the 84 transportation nodes in Section \ref{sec:transnet}. For clustered bus stations, we connect them to a single node in the virtual power network.

To accomplish this, we calculate the geographical distance between transportation nodes using their latitude and longitude coordinates and subsequently define a distance threshold. If the distance between any two nodes within a cluster of transportation nodes is below the threshold, only one node from the cluster will be selected as the corresponding power grid node.

\begin{figure}[!h]
\centering
\includegraphics[width=\linewidth]{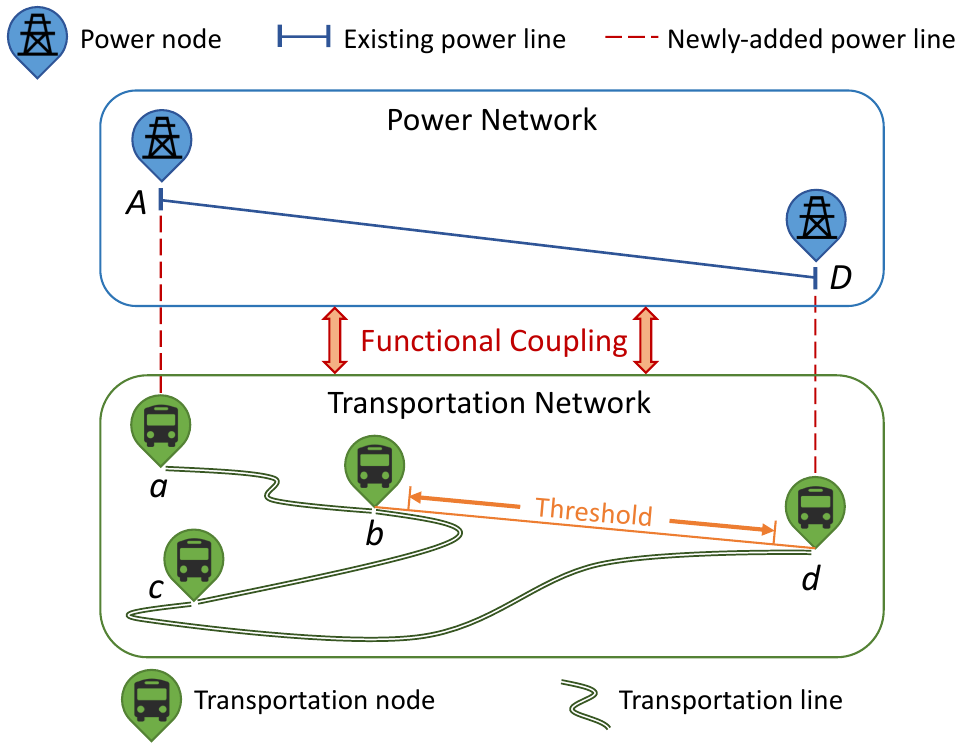}
\caption{Illustration of node selection rules for constructing the virtual power network.}
\label{fig5}
\end{figure}  

As illustrated in Figure \ref{fig5}, the pairwise geographical distances between transportation nodes $ab$, $ac$, and $bc$ are below the threshold, while the distances between node $d$ and nodes $a$, $b$, or $c$ exceed the threshold. Accordingly, we choose bus stations $a$ and $d$ as power grid nodes $A$ and $D$ in the power network, respectively. These nodes occupy the same geographical location in both the transportation and power networks.

\begin{algorithm}
\DontPrintSemicolon
\caption{Power Node Selection}\label{alg:powernode}
\KwData{Selected transportation nodes $N^T$}
\KwResult{Selected power nodes $N^G$}
$N^G \gets \emptyset$\;
$N_{\text{visit}} \gets \emptyset$\;
 \For{$m\leftarrow 1$ \KwTo $|N^T|$}{
    \If{$N^T[m] \notin N_{\text{\textnormal{visit}}}$}{
    $N^G \gets N^G \cup \{N^T[m]\}$\;
    \For{$n\leftarrow m+1$ \KwTo $|N^T|$}{
        \If{$N^T[n] \notin N_{\text{\textnormal{visit}}}$}{
        \If{$\mathcal{D}(N^T[m], N^T[n]) < d_\theta^G$}{
        $N_{\text{visit}} \gets N_{\text{visit}} \cup \{N^T[n]\}$\;
        }
        
        }
    }
    }
 }
 \KwRet{$N^G$}
\end{algorithm}

We develop Algorithm \ref{alg:powernode} to select the power nodes from the set of transportation nodes $N^T$, where the distance between any two power nodes is at least the threshold $d_\theta^G$. The algorithm employs an empty list $N_{\text{visit}}$ to collect the transportation nodes that fall within the specified distance threshold from the selected power nodes. These transportation nodes, which are added to $N_{\text{visit}}$, constitute the cluster nodes associated with the selected nodes identified as new power nodes in the transportation network. The function $\mathcal{D}(\cdot,\cdot)$ is used to calculate the geographical distance between two nodes based on their latitude and longitude coordinates.

The algorithm starts by looping through each transportation node in $N^T$. If the current node has already been identified in $N_{\text{visit}}$, the algorithm skips it and moves on to the next node. Otherwise, it adds the current node to the power node set $N^G$ in line 5 and loops through each subsequent node.

For each subsequent node, the algorithm checks if the distance between the current selected power node and the subsequent node is less than the threshold $d_\theta^G$ in line 8. If it is, the algorithm marks the subsequent node as visited by adding it to $N_{\text{visit}}$ in line 9. If it is not, the algorithm continues to the next subsequent node. In this way, the algorithm can obtain the set of power nodes $N^G$.

To plot the topology of the power network, we need to establish the branches between the selected power nodes. Since the distribution power network typically operates in a radial topology, we aim to create a radial topology of the power network while minimizing the total length of power lines to reduce costs. To achieve this, we use the minimum spanning tree (MST) algorithm, such as Prim's algorithm \cite{prim1957shortest} or Kruskal's algorithm \cite{kruskal1956shortest}, to determine the links between power nodes.

We begin by constructing a graph with the selected power nodes, where each node represents a power node, and the edges between nodes represent potential power lines. The geographical distance between all pairs of nodes is calculated and added as an edge weight to the graph. We then apply the MST algorithm to the graph to find the minimum spanning tree, which identifies the subset of edges that connect all the power nodes with the lowest possible total edge weight. Finally, we visualize the power network topology by plotting the graph and the minimum spanning tree. The NetworkX library in Python is used to implement the above steps and the pseudocode is omitted for brevity.

\begin{figure}[!h]
\centering
\includegraphics[width=\linewidth]{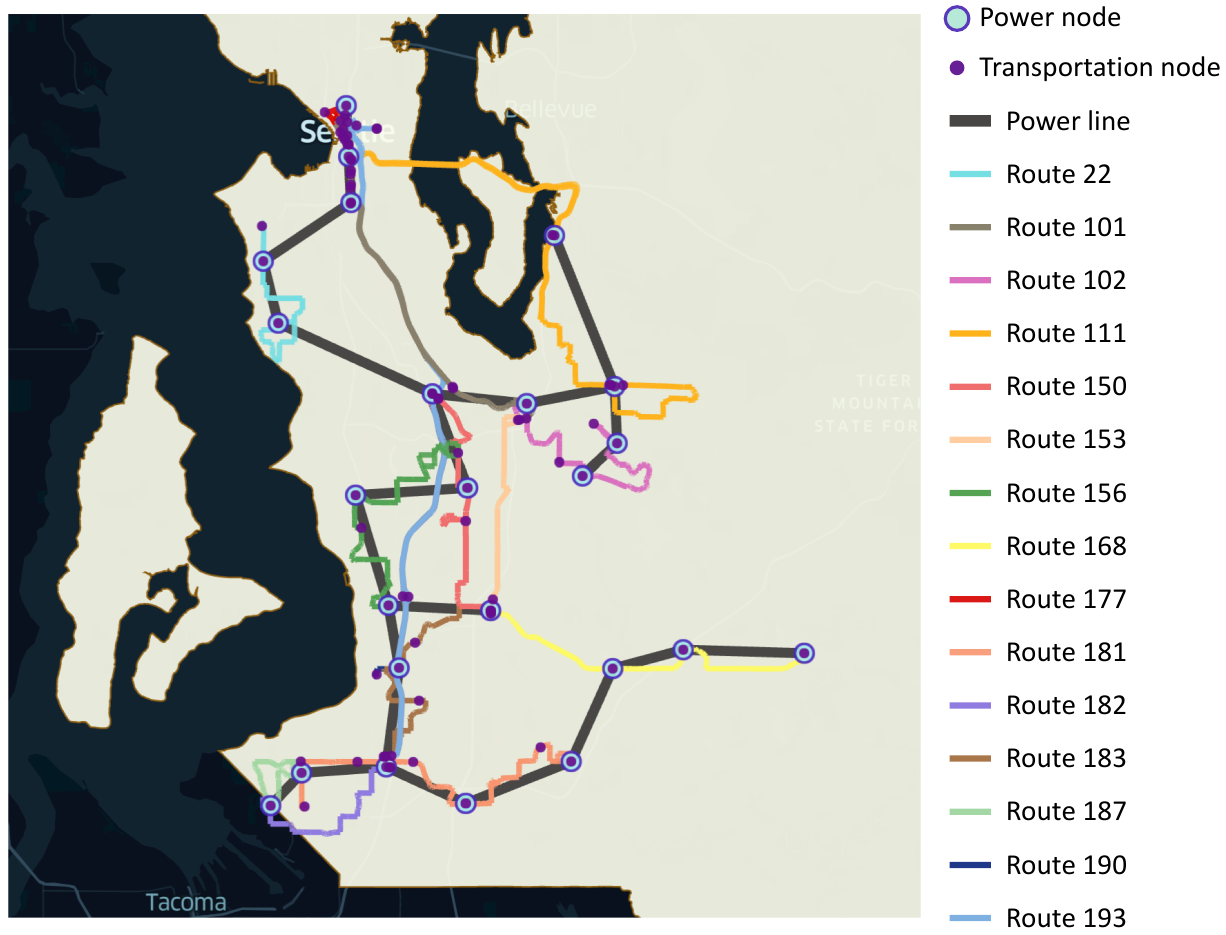}
\caption{Coupled power and transportation network in South King County.}
\label{fig6}
\end{figure} 

As shown in Figure \ref{fig6}, there is a geographic overlap between the nodes of the power network and a subset of nodes in the transportation network, thereby forming the coupled network. Notably, by setting the geographical distance threshold $d_\theta^G$ to 2 km in Algorithm \ref{alg:powernode}, we have identified 24 power grid nodes from the initial pool of 84 transportation nodes. The node selection rules and algorithms utilized in the design of the transportation network and virtual power grid can be extended to other bus systems, highlighting the versatility and applicability of this framework. For ease of access and further exploration, we have made the complete implementation code of our coupled network framework available on our GitHub repository \cite{Xinyi-gtfs_transpowernet}.

\section{Case Studies}\label{sec5}
In this section, we will execute the planning model presented in Section \ref{sec3} on the coupled networks established in Section \ref{sec4}. Initially, we will run the planning model without incorporating any fairness constraints to examine the variations in planning metrics based on various initial SOC levels for BEBs. Subsequently, we will rerun the planning model with fairness considerations to ensure a high degree of horizontal or vertical equity in the planning outcomes. For equity analysis, the census tracts traversed by all 15 bus routes will be partitioned into distinct subareas based on different socio-demographic characteristics.

\subsection{Parameter Settings}
\begin{figure}[!h]
\centering
\includegraphics[width=\linewidth]{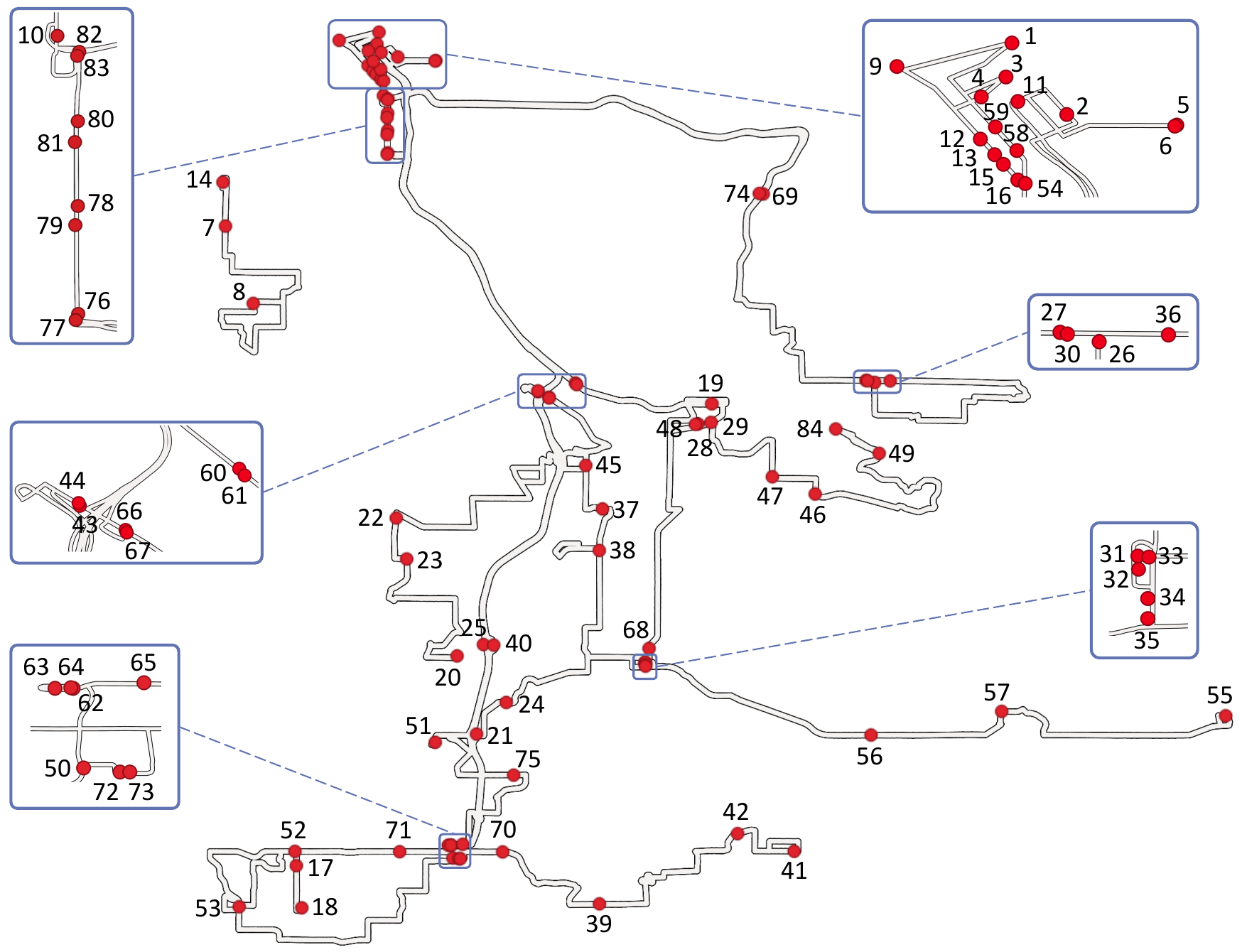}
\caption{Representation of network topology for the 84-Node transportation network.}
\label{fig7}
\end{figure}

\begin{figure}[!h]
\centering
\includegraphics[width=\linewidth]{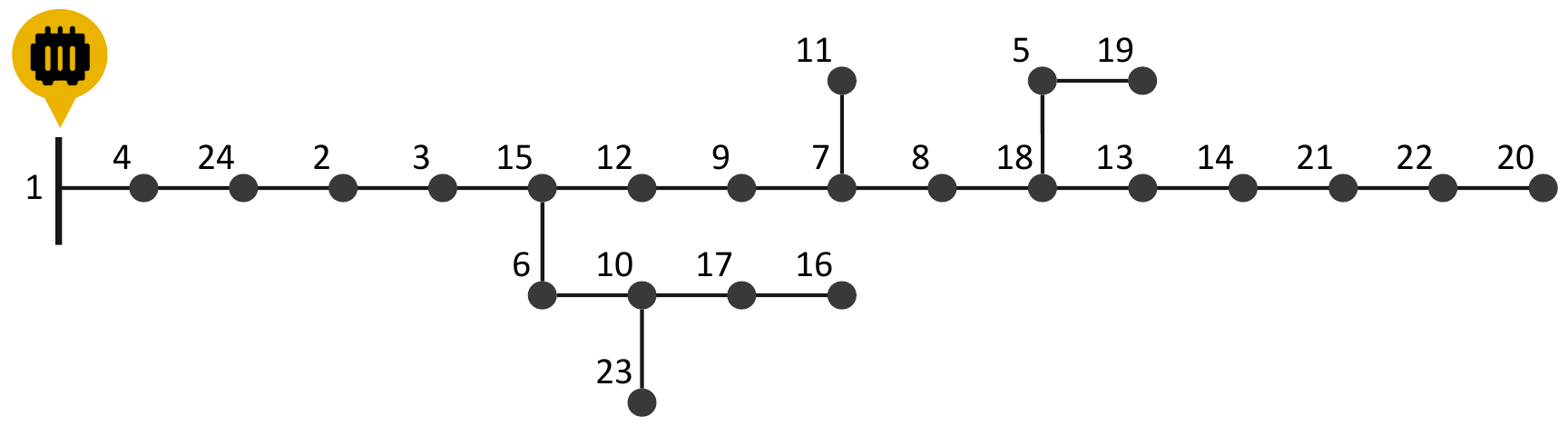}
\caption{Representation of network topology for the 24-node distribution power network.}
\label{fig8}
\end{figure}

The topology of the 84-node transportation network is illustrated in Figure \ref{fig7} and the diagram of the 110 kV high voltage distribution network is shown in Figure \ref{fig8}. For the parameters of the distribution network, please refer to \cite{xinyi-0724_kingcounty_24_node_virtual_power_grid_2023}. The resistance and reactance of each power line reflect the actual geographical distance between its connecting power nodes. And the coupling relationship between the power and transportation network nodes is summarized in Table \ref{tab:couple_relation}. 

\begin{table}[htbp]
  \centering
   \aboverulesep=0ex 
   \belowrulesep=0ex 
  \caption{Geographically Overlapping Nodes between the Transportation Network and the Distribution Power Network}
    \resizebox{\linewidth}{!}{%
    \begin{tabular}{cccc|cccc}
    \toprule
    \multicolumn{1}{p{1.3em}}{Power Node} & \multicolumn{1}{p{1.3em}}{Trans Node} & Latitude & Longitude & \multicolumn{1}{p{1.3em}}{Power Node} & \multicolumn{1}{p{1.3em}}{Trans Node} & Latitude & Longitude \\
    \midrule
    1     & 1     & 47.6167107 & -122.3306 & 13    & 39    & 47.2959099 & -122.24944 \\
    2     & 7     & 47.545311 & -122.38711 & 14    & 41    & 47.315258 & -122.17787 \\
    3     & 8     & 47.5168533 & -122.3769 & 15    & 43    & 47.4843712 & -122.27198 \\
    4     & 10    & 47.5933418 & -122.32896 & 16    & 46    & 47.4468002 & -122.17017 \\
    5     & 17    & 47.3099785 & -122.36103 & 17    & 49    & 47.4616432 & -122.14659 \\
    6     & 19    & 47.4798775 & -122.20813 & 18    & 50    & 47.3127861 & -122.30338 \\
    7     & 20    & 47.3871994 & -122.30184 & 19    & 53    & 47.2948532 & -122.38205 \\
    8     & 21    & 47.3584251 & -122.29468 & 20    & 55    & 47.3651619 & -122.01903 \\
    9     & 22    & 47.4379692 & -122.32423 & 21    & 56    & 47.358078 & -122.14958 \\
    10    & 26    & 47.4877625 & -122.14824 & 22    & 57    & 47.3667755 & -122.10149 \\
    11    & 31    & 47.3848267 & -122.2327 & 23    & 69    & 47.5571404 & -122.18928 \\
    12    & 37    & 47.4413147 & -122.24831 & 24    & 76    & 47.5722656 & -122.32739 \\
    \bottomrule
    \end{tabular} }
  \label{tab:couple_relation}%
\end{table}%

In the objective function \eqref{obj}, the fixed cost of constructing each charging station is represented by $f_{s,m}$ and assumed to be \$200,000 at all bus stations. The fixed cost of building each pile is denoted by $f_{c,m}$ and set at \$25,000. Overhead power line costs are estimated to be \$390,000 per mile. The construction cost of power lines $c_{i,m}$ is determined based on the geographical distance between the power node $i$ and bus station $m$, using the unit cost mentioned above. For the planning period, the power loss time $T$ is assumed to be 15 hours per day. Furthermore, the electricity cost for BEBs is taken to be \$0.20/kWh, which is based on the average rate paid by King County Metro for electricity \cite{eudy2017king}.

To guarantee the efficient performance of BEBs, we have established both lower and upper limits for the SOC of the bus batteries at 10\% and 90\%, respectively. In our planning problem, we consider two types of coaches for BEBs: 40-ft and 60-ft. The coach type for each bus route is provided in Appendix C of the King County Transit report \cite{king_county_metro_king_2020}. We specifically utilize the 40-ft BYD K9M and 60-ft BYD K11M BEB models, which have battery capacities of 313 kWh and 578 kWh, respectively. 
The estimated average energy consumptions per kilometer for 40-ft and 60-ft buses are 1.99 kWh/mile and 3.74 kWh/mile, as indicated in the sources \cite{BYD_K9, BYD_K11M}. According to the official published specifications from BYD \cite{BYD_K9M_intro, BYD_K11M_intro}, the nominal charging powers for K9M and K11M electric buses are 150 kW and 200 kW, respectively. Moreover, we have set a maximum waiting time of 12 minutes at each station to ensure that BEBs receive timely energy support.

As a benchmark for the model scale, we employed Gurobi \cite{gurobi} to solve the MISOCP planning problem on a laptop equipped with a 2.4 GHz Quad-Core Intel Core i5 processor and 8 GB of memory. Using a tolerance of $1.00e^{-4}$, the model can be solved within one minute.

\subsection{Fairness Zone Division}
Jain's index, as defined in \eqref{eq:jain}, serves as a metric for assessing transit equity across multiple areas. To evaluate fairness in the BEB planning area of South King County with Jain's index, we need to partition the region into several distinct areas. Notably, census tracts, which offer stable and relatively permanent geographic units for statistical data presentation \cite{bureau_glossary}, are well-suited for this purpose. Previous studies on transit equity have conducted analyses at the census tract level \cite{ju2020equity, ferenchak2019equity}, considering the social demographic characteristics associated with each tract. Therefore, we will evaluate the fairness of our planning outcome based on census tracts. However, considering the small size of individual census tracts, they may not contain a sufficient number of bus routes for meaningful equity analysis. To address this issue, we will merge census tracts into larger subareas based on two specific criteria related to different dimensions of transit equity.

\subsubsection{Population-Based Census Tract Merging}\label{population_ct}
\begin{figure}[t]
\centering
\includegraphics[width=\linewidth]{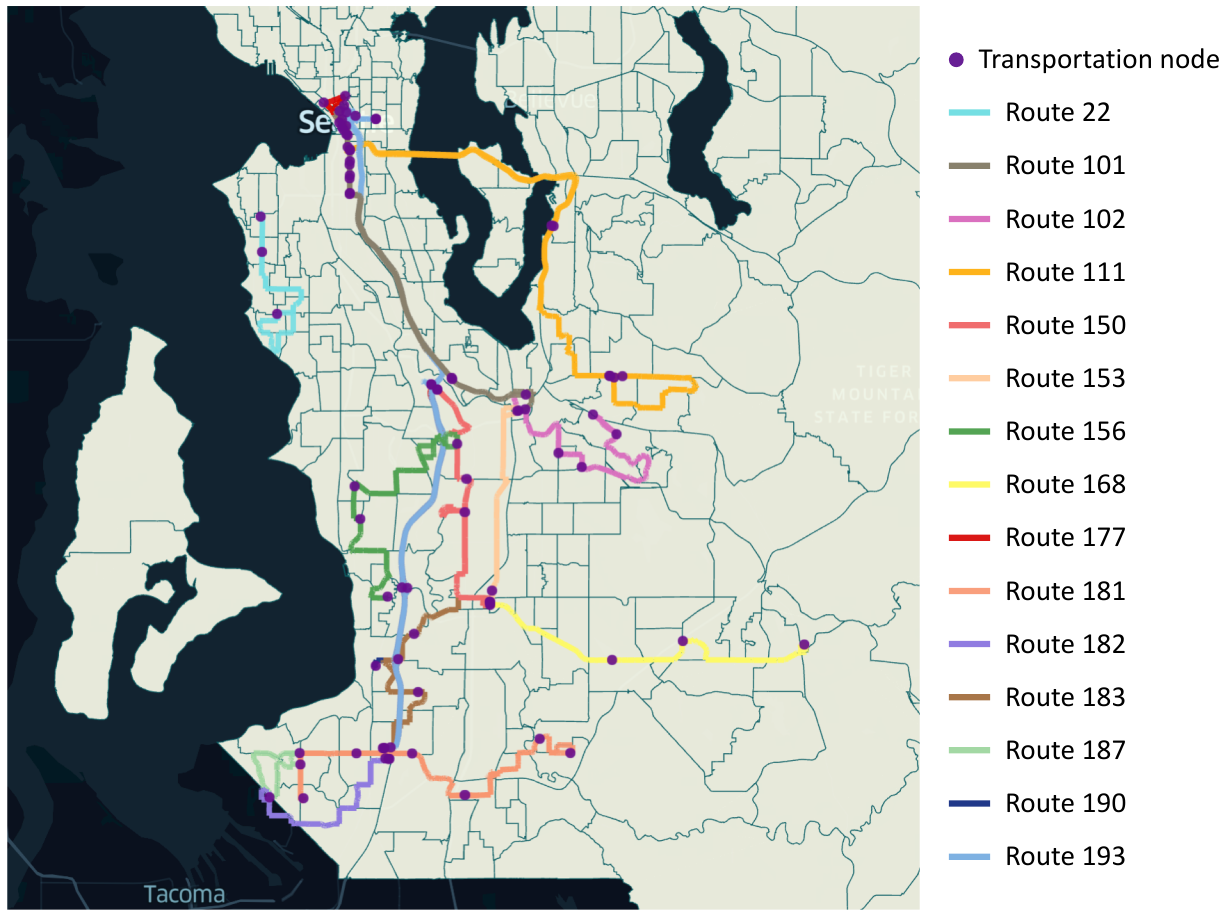}
\caption{Illustration of the census tract map in South King County.}
\label{fig9}
\end{figure}

The map of the census tract polygons, obtained from King County geographic information system open data \cite{census_tract_gis}, is presented in Figure \ref{fig9}. Our focus lies primarily on census tracts that are intersected by bus routes, as the residents in these areas are more vulnerable to the impacts of air pollution and the bus services associated with these routes.

To prioritize the development of horizontally equitable bus services, we aim to merge these census tracts into larger areas with comparable total populations and an adequate number of bus routes. By doing so, the fairness constraint \eqref{cons:fair} will encourage a balanced distribution of BEB resources across the merged subareas, ensuring that the BEB route ratio ($w_h$) for each resident in any of these areas is similar to that of residents in other areas. This approach allows us to analyze horizontal equity in the BEB planning outcomes by merging census tracts into larger subareas based on their respective populations.

\begin{figure}[!h]
\centering
\includegraphics[width=\linewidth]{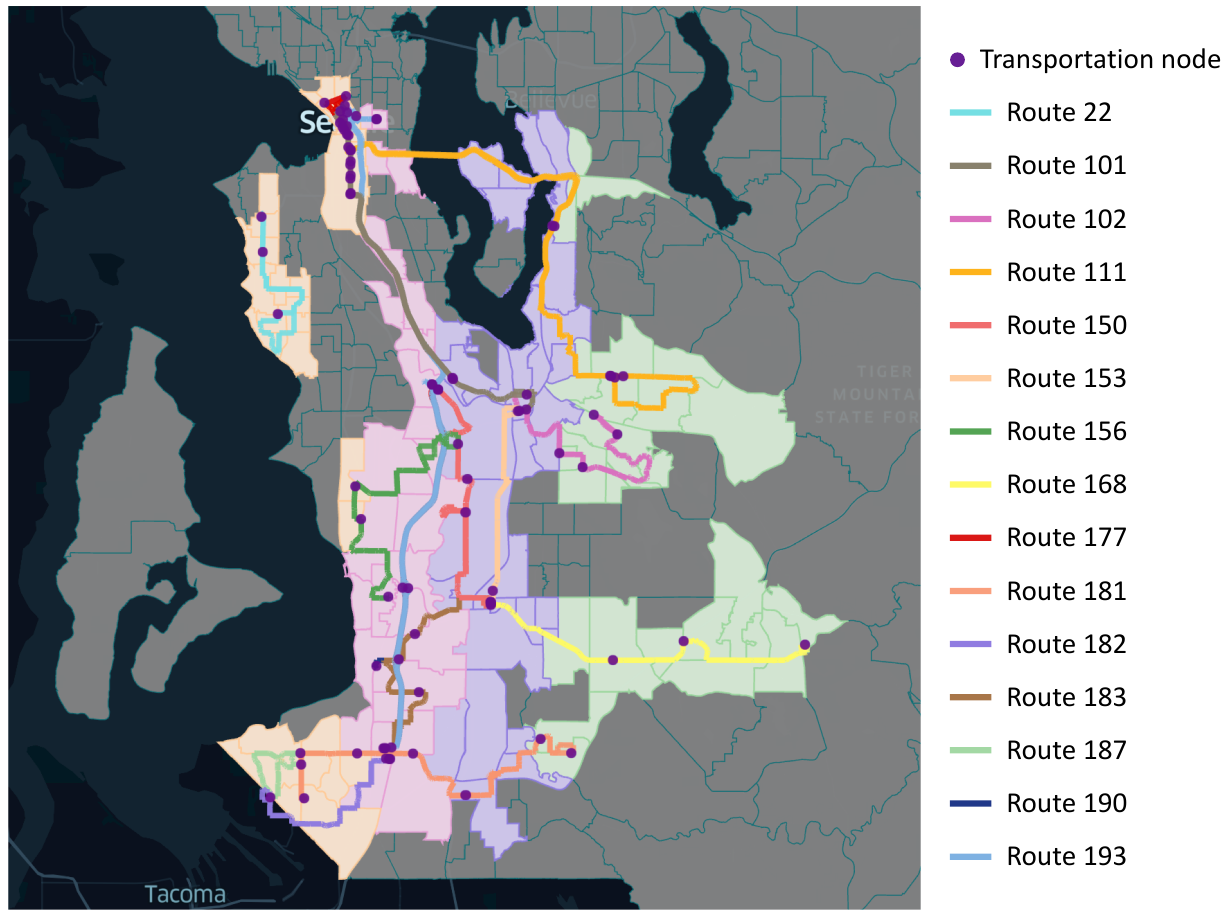}
\caption{Map of 4 subareas formed by aggregating census tracts based on the population feature.}
\label{fig10}
\end{figure} 
As the majority of the bus routes in this region run from north to south, we have horizontally arranged the four merged census tracts, as depicted in Figure \ref{fig10}. We obtained the population data for each tract from the consolidated demographics index for King County census tracts \cite{KingCounty_demographicindex}. The resulting merged areas have been designated as Zone 1, Zone 2, Zone 3, and Zone 4, ordered from left to right. To achieve a balanced population distribution across the merged areas, Zone 1, Zone 2, Zone 3, and Zone 4 have respective populations of 173,501, 179,359, 176,994, and 165,661 people. The bus routes passing through each subarea are as follows: Zone 1 includes routes 22, 101, 102, 111, 150, 156, 177, 181, 187, 190, and 193. Zone 2 includes routes 156, 177, 181, 182, 183, 187, 190, and 193. Zone 3 includes routes 101, 102, 150, 153, 156, 168, 181, 183, and 193. Zone 4 includes routes 102, 111, 168, and 181.

\subsubsection{Bus-Commuter-Based Census Tract Merging}\label{bus_commuter_division}
In contrast to the population-based method in Section \ref{population_ct}, the evaluation of vertically equitable bus services focuses on a specific community that is particularly vulnerable to bus services: bus commuters. Rather than considering the benefits of BEBs for all residents, we merge census tracts into subareas based on the total number of workers who rely on buses for their daily commute. By creating subareas with similar bus-commuter populations, we aim to distribute the benefits of BEBs more evenly among this specific group. To gather the necessary information, we refer to the US census American community survey data table for the "journey to work" subject area \cite{KingCounty_acs}. Identifying the bus-commuter community as a group heavily reliant on bus services, we incorporate their specific needs into our vertical analysis by enforcing the fairness constraint \eqref{cons:fair} to ensure that all bus commuters in these subareas have equitable access to BEB routes.

\begin{figure}[!h]
\centering
\includegraphics[width=\linewidth]{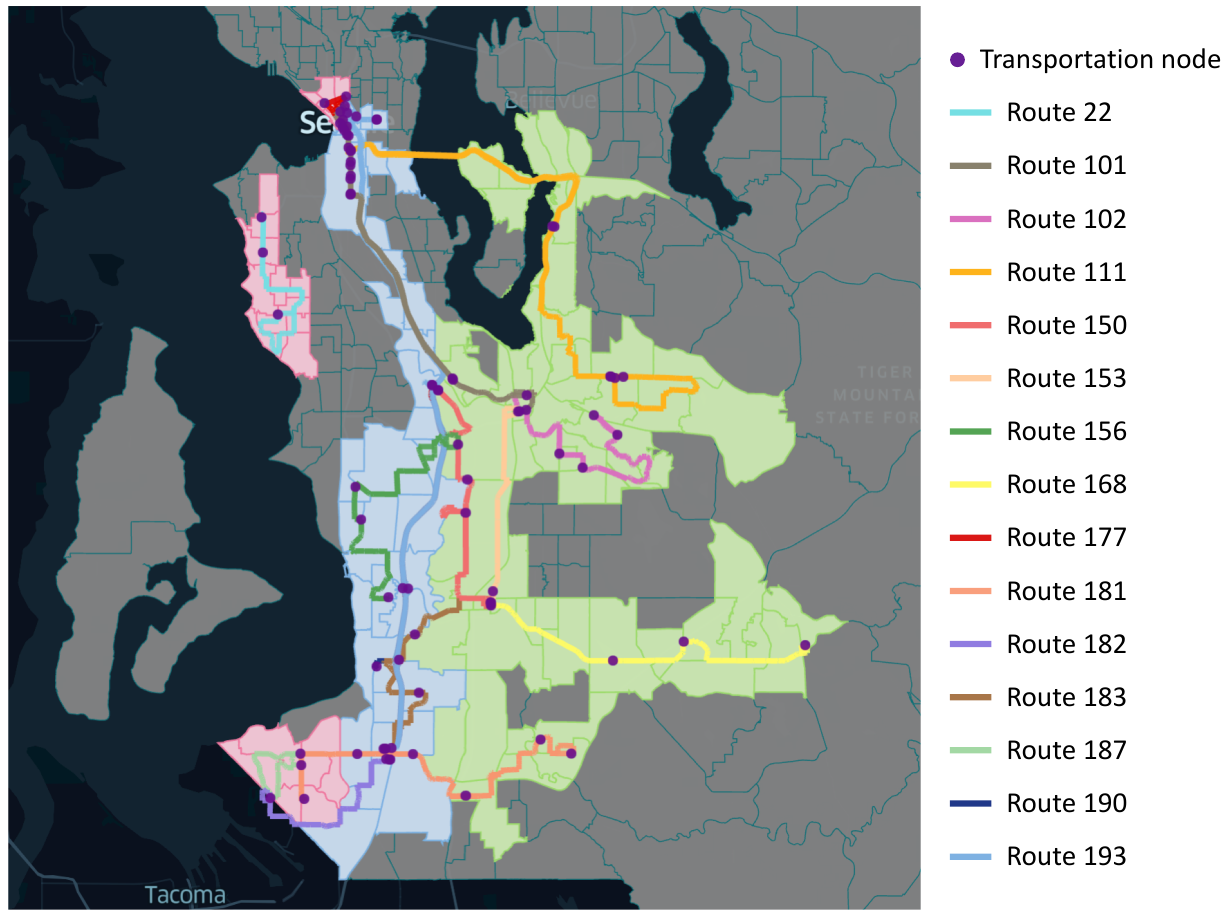}
\caption{Map of 3 subareas formed by aggregating census tracts based on the bus-commuter feature.}
\label{fig11}
\end{figure} 

We have merged the census tracts into three subareas, ensuring an equitable distribution of the bus-commuter population among each subarea, as illustrated in Figure \ref{fig11}. These subareas are named Region 1, Region 2, and Region 3, arranged from left to right. The Seattle downtown area, located in the top left corner of the figure, contains a significant concentration of bus commuters. As a result, Region 1, although relatively smaller in size, has a similar number of bus commuters compared to the other two regions. Specifically, Region 1, Region 2, and Region 3 have 11,658, 11,985, and 11,526 bus commuters.

In Region 1, the following 10 bus routes pass through: 22, 101, 102, 111, 150, 177, 181, 187, 190, and 193. Region 2 comprises 12 routes: 101, 102, 111, 150, 156, 177, 181, 182, 183, 187, 190, and 193. Region 3 consists of 12 bus routes: 101, 102, 111, 150, 153, 156, 168, 177, 181, 183, 190, and 193. Each subarea has its unique set of bus routes, with Region 1 featuring route 22, Region 2 having route 182, and Region 3 having routes 153 and 168 that are not found in the other subareas.

\subsection{Planning Results without Fairness}
Assuming a 10-year planning period, we begin by solving the planning model \eqref{planning_wo_fair} without incorporating any fairness constraints. Table \ref{tab:chargetimes} provides information on the frequency of on-route charging required to sustain the round trips for all 15 bus routes. The table illustrates that as the initial SOC of the BEB batteries increases, the overall number of required charging sessions decreases. Notably, once the initial SOC exceeds 50\%, all BEBs are capable of completing their round trips without the need for on-route charging. This finding highlights the importance of ensuring that BEBs are charged to adequate SOC levels prior to departure, which can help to optimize their operational efficiency and minimize the need for additional charging infrastructure.

\begin{table}[htbp]
  \centering
  \caption{On-Route Charging Frequency per Round-Trip for Each Bus Route at Different Departure SOC $\theta_0$}
  \resizebox{\linewidth}{!}{
    \begin{tabular}{cccccccc}
    \toprule   
    \multirow{2}[4]{*}{Route} & \multicolumn{1}{c}{\multirow{2}[4]{*}{\makecell{Round-trip\\Length (mile)}}} & \multicolumn{1}{c}{\multirow{2}[4]{*}{\makecell{Charge \\Power (kW)}}} & \multicolumn{5}{c}{Initial SOC $\theta_0$} \\
\cmidrule{4-8}     &     &       &  $0.1$   & $0.2$  & $0.3$   & $0.4$   & $\geq 0.5$  \\
    \midrule
    22    & 13.82 & 150   & 1     & 0     & 0     & 0     & 0 \\
    101   & 28.28 & 200   & 3     & 2     & 0     & 0     & 0 \\
    102   & 47.81 & 200   & 5     & 4     & 2     & 1     & 0 \\
    111   & 52.34 & 200   & 5     & 4     & 3     & 1     & 0 \\
    150   & 44.18 & 200   & 5     & 3     & 2     & 0     & 0 \\
    153   & 16.37 & 150   & 2     & 1     & 0     & 0     & 0 \\
    156   & 25.04 & 150   & 2     & 1     & 0     & 0     & 0 \\
    168   & 24.35 & 150   & 2     & 1     & 0     & 0     & 0 \\
    177   & 48.66 & 200   & 5     & 4     & 2     & 1     & 0 \\
    181   & 30.08 & 150   & 2     & 1     & 0     & 0     & 0 \\
    182   & 15.10 & 150   & 1     & 0     & 0     & 0     & 0 \\
    183   & 21.79 & 150   & 2     & 1     & 0     & 0     & 0 \\
    187   & 11.81 & 150   & 1     & 0     & 0     & 0     & 0 \\
    190   & 41.79 & 200   & 4     & 3     & 2     & 0     & 0 \\
    193   & 50.62 & 200   & 5     & 4     & 2     & 1     & 0 \\
    \bottomrule
    \end{tabular}
    }
  \label{tab:chargetimes}%
\end{table}%

\begin{table}[!h]
  \centering
  \caption{Summary of Planning Results without Fairness Consideration When Initial SOC $\theta_0$ Ranges from 0.1 to 0.4}
  \resizebox{\linewidth}{!}{%
    \begin{tabular}{lcccc}
    \toprule
    Planning Metric  & $\theta_0 = 0.1$   & $\theta_0 = 0.2$  & $\theta_0 = 0.3$   & $\theta_0 = 0.4$    \\
    \midrule
    Number of stations & 27    & 14    & 7     & 3      \\
    Number of piles & 45    & 29    & 13    & 4      \\
    Total cost  & \$10,118,392 & \$4,329,727 & \$2,033,864 & \$926,386  \\
    Station investment & \$5,400,000 & \$2,800,000 & \$1,400,000 & \$600,000  \\
    Pile investment & \$1,125,000 & \$725,000 & \$325,000 & \$100,000 \\
    Power line investment & \$3,156,620 & \$463,221 & \$62,508 & \$0 \\
    Power loss cost & \$436,772 & \$341,506 & \$246,356 & \$226,386 \\
    \bottomrule
    \end{tabular}%
    }
  \label{tab:plan_wo_fair}
\end{table}%

Table \ref{tab:plan_wo_fair} presents the planning results for initial SOC values ranging from 0.1 to 0.4, considering that on-route charging is no longer needed for the 15 BEBs when their initial SOC reaches or exceeds 50\%. The table demonstrates that increasing the initial SOC leads to a decrease in the total planning cost. This reduction can be attributed to the decreased charging demand resulting from a higher initial SOC, which in turn reduces the investment required for charging stations and piles. Additionally, the cost of power line investment decreases consistently as the length of power lines depends on the number of charging stations and the distance between the stations and the power nodes in which they are integrated. When $\theta_0=0.4$, the power line investment becomes zero because the three charging stations are built directly on the power nodes, eliminating the need for extra power lines to connect the stations to the power grid. Moreover, the cost of power loss declines steadily with increasing $\theta_0$, reflecting the fact that BEBs with adequate SOC require less electric power, resulting in lower current flow in the power lines and reduced power loss.

\begin{figure}[!h]
\centering
\includegraphics[width=\linewidth]{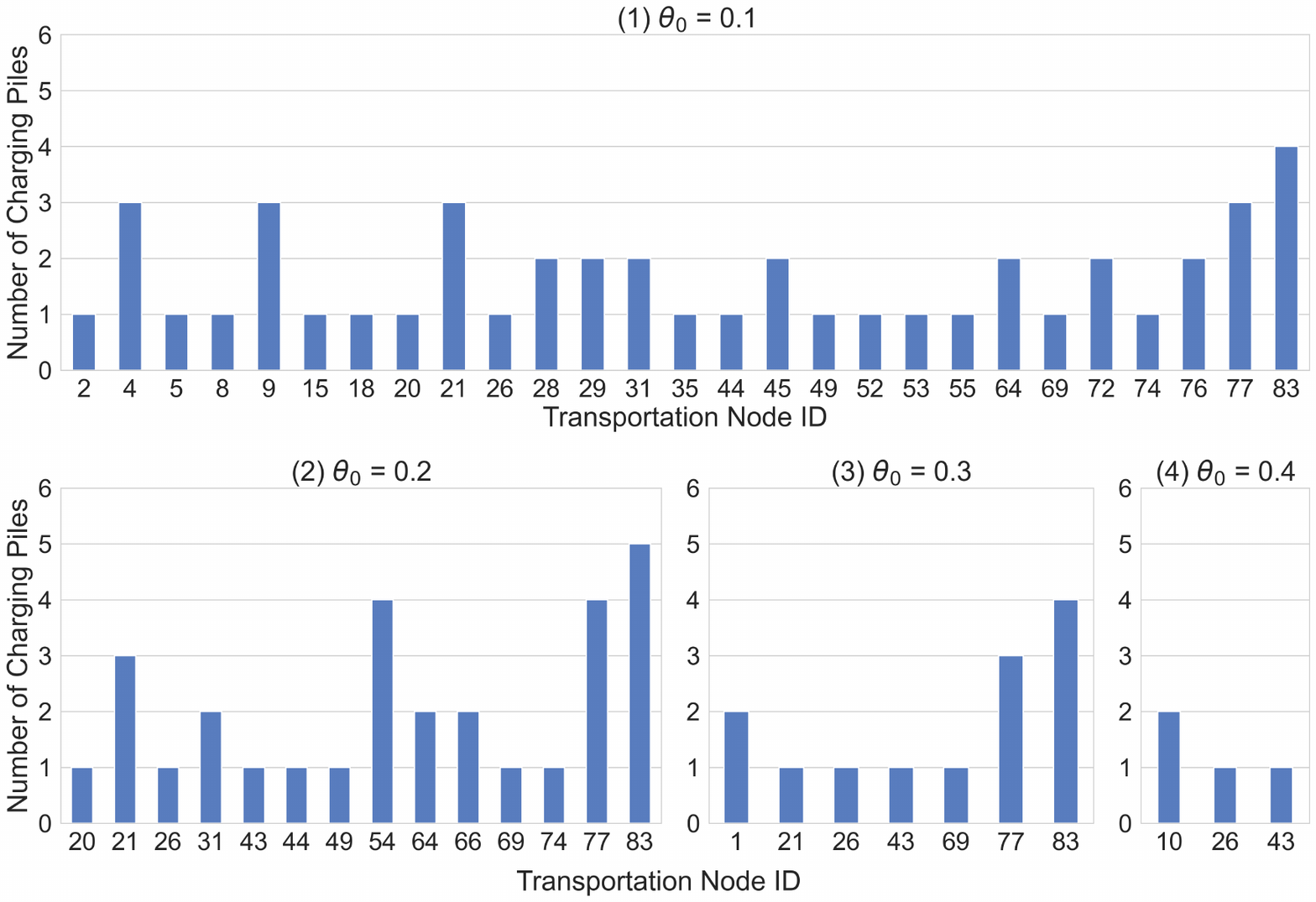}
\caption{Optimal placement and charging pile allocation result of charging stations without fairness consideration.}
\label{fig12}
\end{figure}

Figure \ref{fig12} displays the siting and sizing outcomes of on-route fast charging stations, represented by the transportation node ID and the number of charging piles installed at each station. Since each bus route has its dedicated charging piles, we can determine the number of bus routes being charged at each station by counting the corresponding charging piles. When $\theta_0 = 0.1$, all origin stations of the 15 identified BEBs are included in these 27 charging stations. Notably, transportation node 83 situated in the Industrial District (SODO Busway \& S Royal Brougham Way) hosts the highest number of BEB routes that charge here, totaling 4 routes. This node serves as an on-route bus stop for 5 BEB routes, which is consistent with selection rule 1) that designates common stops. Additionally, among the nodes with three charging piles, nodes 4 and 9 are origin stations for two bus routes each, while nodes 21 and 77 serve for no fewer than 3 bus routes.

When $\theta_0 = 0.2$, only one origin station for route 153, located at node 31, still requires the construction of charging stations. However, as $\theta_0$ increases to $0.3$ and $0.4$, none of the origin stations require charging stations since the BEBs have enough energy to run the first few stops while maintaining a safe SOC. With larger initial SOC, the number of charging stations decrease evidently which aligns with the findings in Table \ref{tab:plan_wo_fair}. From $\theta_0 = 0.2$ to $0.4$, the number of BEB routes requiring on-route charging decreases. At $\theta_0 = 0.4$, only four routes require on-route charging, as confirmed by the data in Table \ref{tab:chargetimes}. The nodes with the most charging piles built between $\theta_0 = 0.2$ and $0.4$ are common stops, including nodes 54, 77, 83, and 10. This highlights the importance of building on-route charging stations at stops that serve multiple routes and further validates the effectiveness of selection rule 1) in forming the coupled network.

\subsection{Planning Results with Fairness Consideration}
In this section, we will maintain the assumption of a 10-year planning period. However, we will now incorporate the fairness measurement \eqref{cons:fair_socp} into the planning model, as represented by \eqref{planning_with_fair}. Notably, a maximum of 5 BEB routes ($I_{\text{max}} = 5$), which is one-third of the total bus routes, will be selected for investment. To include all 15 bus routes as candidate BEB routes, we will set the initial SOC to 0.1 based on the information provided in Table \ref{tab:chargetimes}. This approach will enable us to evaluate both the horizontal equity of the BEB route ratio across the population-based merged subareas and the vertical equity within the bus-commuter-based merged subareas.

\begin{table}[!h]
  \centering
  \caption{Summary of Planning Results Considering Horizontal Equity with a Maximum Number of BEB Routes $I_{\text{max}} = 5$}
   \resizebox{\linewidth}{!}{
    \begin{tabular}{p{9em}p{4.5em}<{\centering}p{4.5em}<{\centering}p{4.5em}<{\centering}p{4.5em}<{\centering}}
    \toprule
    Planning Metric & \multicolumn{1}{c}{$\eta=0$} & \multicolumn{1}{c}{$\eta=0.9$} & \multicolumn{1}{c}{$\eta=0.95$} & \multicolumn{1}{c}{$\eta=0.99$} \\
    \midrule
    Number of stations & \multicolumn{1}{c}{7} & \multicolumn{1}{c}{7} & \multicolumn{1}{c}{7} & \multicolumn{1}{c}{10} \\
    Number of piles & \multicolumn{1}{c}{8} & \multicolumn{1}{c}{8} & \multicolumn{1}{c}{8} & \multicolumn{1}{c}{11} \\
    Total cost & \multicolumn{1}{c}{\$2,002,226} & \multicolumn{1}{c}{\$2,002,226} & \multicolumn{1}{c}{\$2,166,931} & \multicolumn{1}{c}{\$2,879,586} \\
    Station investment & \multicolumn{1}{c}{\$1,400,000} & \multicolumn{1}{c}{\$1,400,000} & \multicolumn{1}{c}{\$1,400,000} & \multicolumn{1}{c}{\$2,000,000} \\
    Pile investment & \multicolumn{1}{c}{\$200,000} & \multicolumn{1}{c}{\$200,000} & \multicolumn{1}{c}{\$200,000} & \multicolumn{1}{c}{\$275,000} \\
    Power line investment & \multicolumn{1}{c}{\$142,253} & \multicolumn{1}{c}{\$142,253} & \multicolumn{1}{c}{\$307,758} & \multicolumn{1}{c}{\$340,780} \\
    Power loss cost & \multicolumn{1}{c}{\$259,973} & \multicolumn{1}{c}{\$259,973} & \multicolumn{1}{c}{\$259,173} & \multicolumn{1}{c}{\$263,806} \\
    Fairness index & \multicolumn{1}{c}{0.915800} & \multicolumn{1}{c}{0.915800} & \multicolumn{1}{c}{0.959993} & \multicolumn{1}{c}{0.992838} \\
    BEB route ID & 182, 187, 168, 153, 22 & 182, 187, 168, 153, 22 & 187, 168, 183, 153, 22 & 182, 168, 153, 22, 190 \\
    \bottomrule
    \end{tabular}
    }
  \label{tab:fz_res}
\end{table}%

\begin{figure}[!h]
\centering
\includegraphics[width=\linewidth]{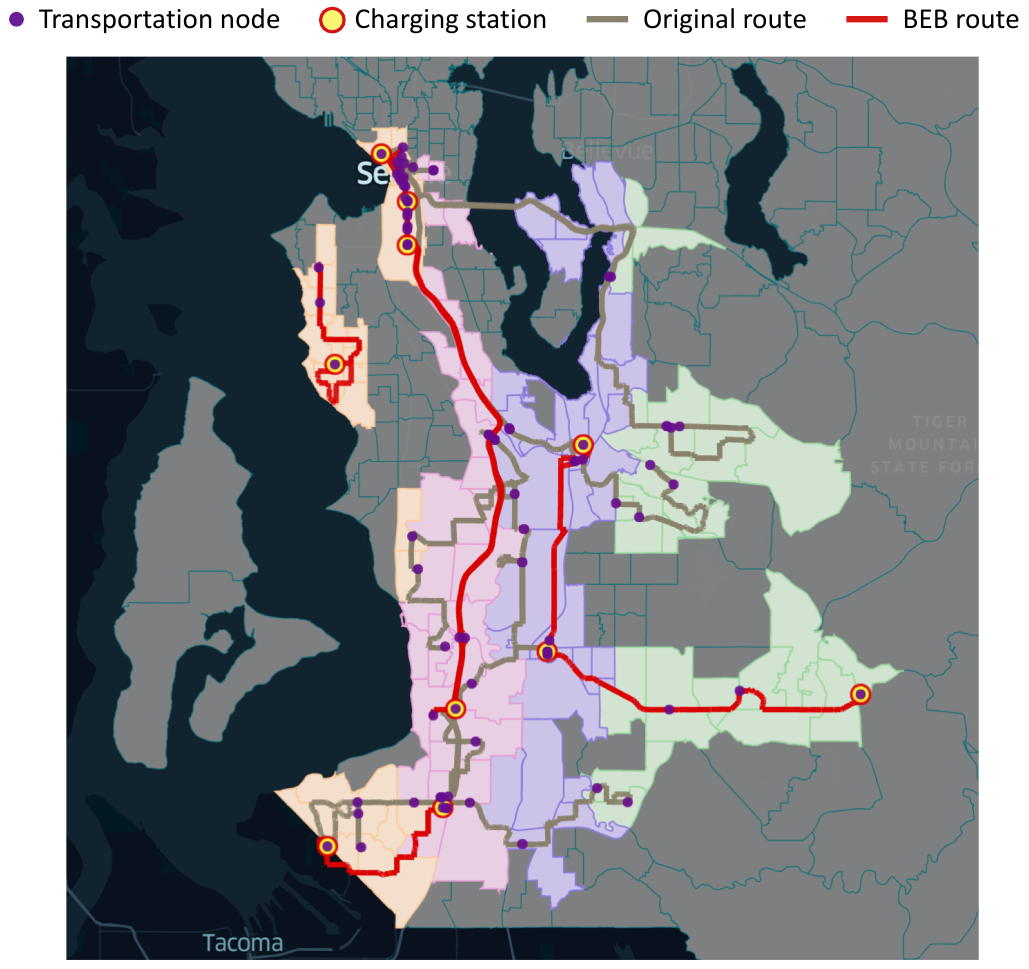}
\caption{Planning results reflecting the highest level of horizontal equity across 4 subareas with $\eta=0.99$.}
\label{fig13}
\end{figure}

Table \ref{tab:fz_res} presents the planning results of the horizontal equity analysis across four subareas merged based on the population feature. The fairness level $\eta$ ranges from $0$ to $0.99$, with a value of $0$ renderings \eqref{cons:fair_socp} invalid. In such cases, the initial fairness index $f(w)$ is computed, prioritizing the minimization of the total planning cost. As the value of $\eta$ increases, a stricter rule is imposed on the equitable distribution of BEB routes among the four subareas. Figure \ref{fig13} illustrates the allocation of these five BEB routes when $\eta = 0.99$, demonstrating a similar proportion of BEB routes to all bus routes in each subarea.

Without considering \eqref{cons:fair_socp}, the initial fairness index achieved through the most cost-effective planning scheme is $0.915800$, surpassing the fairness level of $0.9$. Therefore, we observe that the planning results are identical for fairness levels of $0$ and $0.9$ in Table \ref{tab:fz_res}. However, as the fairness level increases to $0.95$, we observe a corresponding rise in the planning cost, primarily due to increased power line investment expenses. This adjustment is necessary to ensure a higher level of fairness, leading to a reconsideration of the five BEB route IDs. Specifically, from fairness levels of $0.9$ to $0.95$, route 183 replaces route 182 as one of the BEB routes to be invested in.

At a fairness level of $0.99$, the planning model requires three additional charging stations and three more charging piles to accommodate the replacement of routes 187 and 183 with routes 182 and 190 as BEB routes. This expansion of the charging infrastructure leads to increased costs in both power line investments and power loss. The planning model prioritizes fairness by selecting bus routes with higher on-route charging demand to be included as BEB routes, even if it results in higher planning costs. These findings underscore the inherent trade-off between equity and economic efficiency in the planning process. While striving for an equitable distribution of BEB routes, compromises need to be made in terms of increased economic expenses.

Similarly, using the three subareas obtained in Section \ref{bus_commuter_division}, we solve \eqref{planning_with_fair} again and present the planning results in Table \ref{tab:ct_res}. Notably, the planning metrics for $\eta = 0$ in Table \ref{tab:ct_res} and Table \ref{tab:fz_res} are nearly identical, with only slight variations in the calculated fairness index due to the utilization of different subareas. Within the bus-commuter-based merged subareas, the initial fairness index is $0.687317$, which falls below the threshold of $0.9$. Consequently, the planning outcomes for fairness levels of $0$ and $0.9$ are no longer the same.

\begin{table}[htbp]
  \centering
  \caption{Summary of Planning Results Considering Vertical Equity with a Maximum Number of BEB Routes $I_{\text{max}} = 5$}
  \resizebox{\linewidth}{!}{
    \begin{tabular}{p{9em}p{4.5em}<{\centering}p{4.5em}<{\centering}p{4.5em}<{\centering}p{4.5em}<{\centering}}
    \toprule
    Planning Metric & \multicolumn{1}{c}{$\eta=0$} & \multicolumn{1}{c}{$\eta=0.9$} & \multicolumn{1}{c}{$\eta=0.95$} & \multicolumn{1}{c}{$\eta=0.99$} \\
    \midrule
    Number of stations & \multicolumn{1}{c}{7} & \multicolumn{1}{c}{8} & \multicolumn{1}{c}{8} & \multicolumn{1}{c}{11} \\
    Number of piles & \multicolumn{1}{c}{8} & \multicolumn{1}{c}{9} & \multicolumn{1}{c}{9} & \multicolumn{1}{c}{12} \\
    Total cost & \multicolumn{1}{c}{\$2,002,226} & \multicolumn{1}{c}{\$2,403,729} & \multicolumn{1}{c}{\$2,825,122} & \multicolumn{1}{c}{\$2,960,454} \\
    Station investment & \multicolumn{1}{c}{\$1,400,000} & \multicolumn{1}{c}{\$1,600,000} & \multicolumn{1}{c}{\$1,600,000} & \multicolumn{1}{c}{\$2,200,000} \\
    Pile investment & \multicolumn{1}{c}{\$200,000} & \multicolumn{1}{c}{\$225,000} & \multicolumn{1}{c}{\$225,000} & \multicolumn{1}{c}{\$300,000} \\
    Power line investment & \multicolumn{1}{c}{\$142,253} & \multicolumn{1}{c}{\$307,758} & \multicolumn{1}{c}{\$729,021} & \multicolumn{1}{c}{\$182,242} \\
    Power loss cost & \multicolumn{1}{c}{\$259,973} & \multicolumn{1}{c}{\$270,972} & \multicolumn{1}{c}{\$271,101} & \multicolumn{1}{c}{\$278,212} \\
    Fairness index & \multicolumn{1}{c}{0.687317} & \multicolumn{1}{c}{0.901361} & \multicolumn{1}{c}{0.986707} & \multicolumn{1}{c}{0.996923} \\
    BEB route ID &  182, 187, 168, 153, 22 & 182, 187, 168, 183, 153 & 181, 182, 187, 183, 153 & 182, 187, 168, 177, 153 \\
    \bottomrule
    \end{tabular}
    }
  \label{tab:ct_res}%
\end{table}%

As the fairness level increases from $0$ to $0.9$, there is a corresponding increase in the planning cost, and an additional charging station is required when $\eta = 0.9$. This adjustment involves replacing route 22 with route 183. When $\eta$ further increases to $0.95$, the investment cost in charging infrastructure remains relatively stable, but there is a significant rise in power line investment. This change can be attributed to the altered locations of the charging stations. Interestingly, at $\eta = 0.99$, although three additional charging stations must be constructed, there is a reduction in the investment required for power lines. This is due to the decreased total distance between the charging stations and the power grid nodes. However, the increase in both power loss costs and investment in charging infrastructure outweighs the savings achieved, leading to the highest planning cost when $\eta = 0.99$.

\begin{figure}[!h]
\centering
\includegraphics[width=\linewidth]{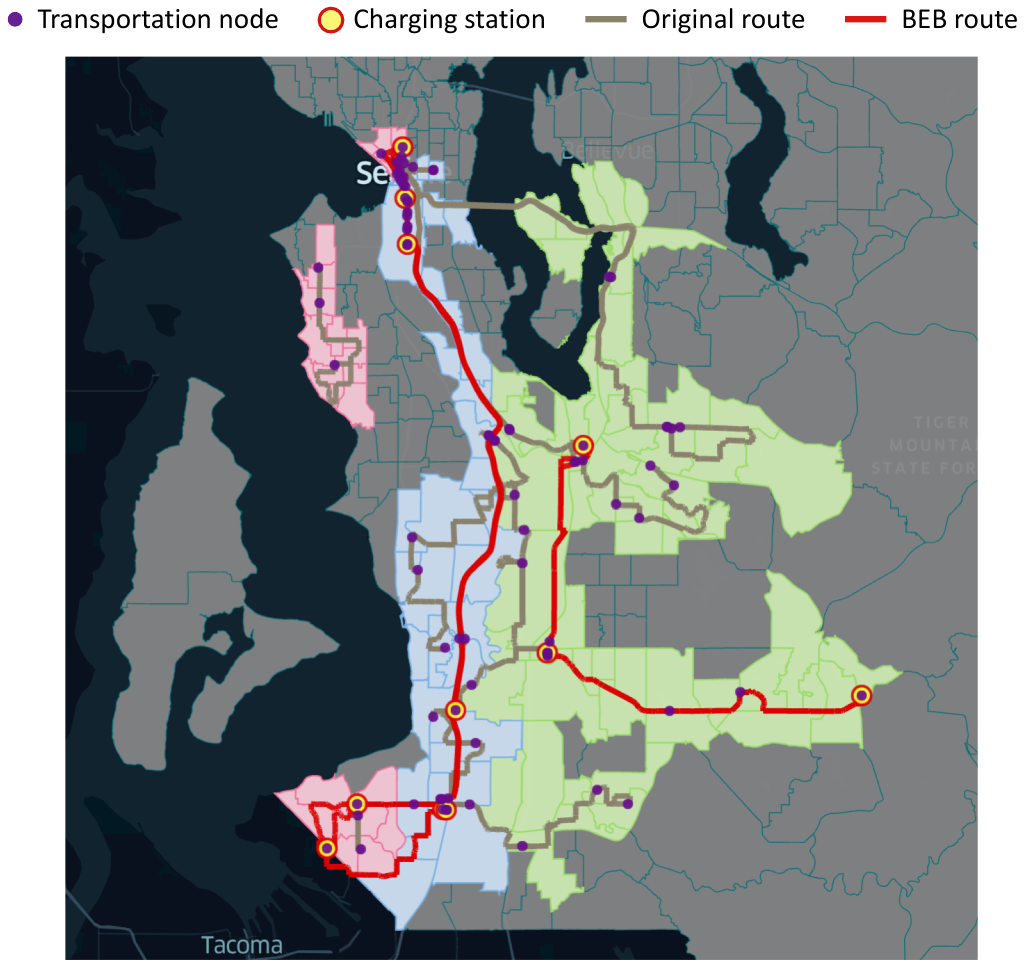}
\caption{Planning results reflecting the highest level of vertical equity across 3 subareas with $\eta=0.99$.}
\label{fig14}
\end{figure}

The distribution of the five BEB routes and the locations of their charging stations within the three bus-commuter-based merged subareas are visualized in Figure \ref{fig14} for a fairness level of $\eta = 0.99$. Comparing this figure with Figure \ref{fig13}, we can observe that routes 22 and 190 from Figure \ref{fig13} have been replaced by routes 187 and 177 in Figure \ref{fig14}. This adjustment from horizontal equity to vertical equity results in longer BEB routes (as indicated in Table \ref{tab:chargetimes}) primarily located in the western portion of the census tracts.

This observation suggests that residents in the western region have a higher reliance on bus transportation, which aligns with the actual transportation landscape. In contrast, the eastern part of King County shows a scarcity of bus routes, indicating that residents in this area must rely on alternative transportation methods, such as household cars, to fulfill their commuting needs.

\begin{figure*}[htbp]
\centering
\includegraphics[width=\textwidth]{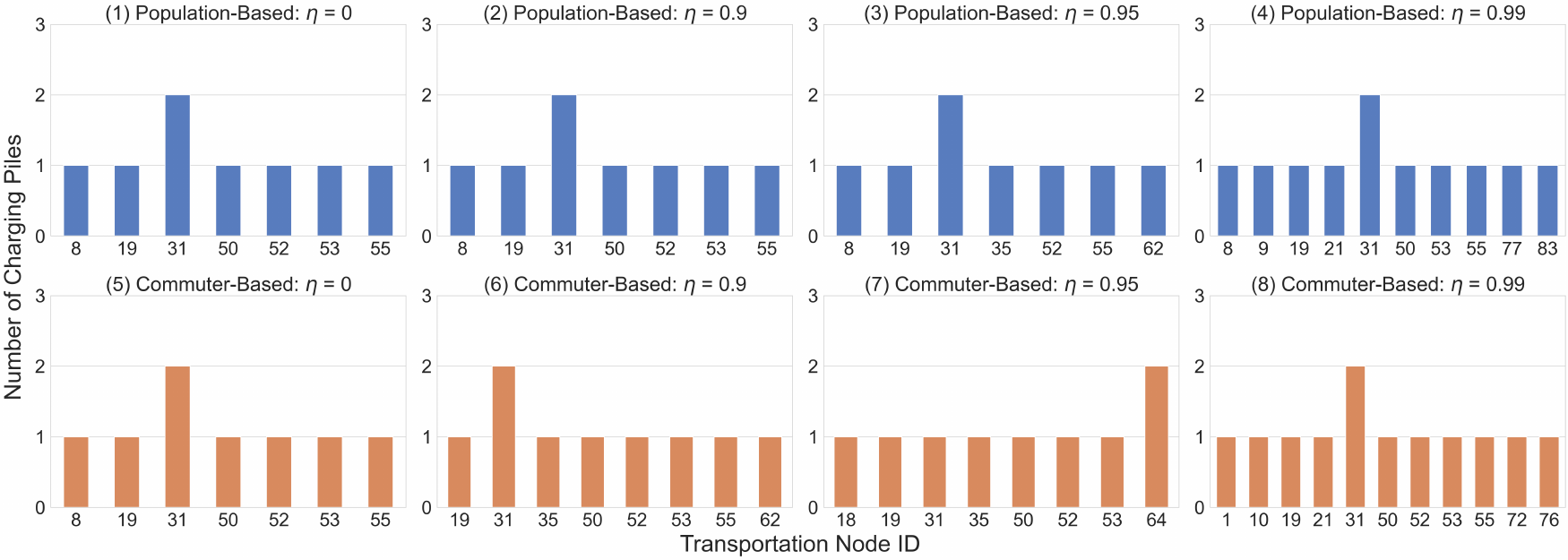}
\caption{Optimal placement and charging pile allocation result of charging stations considering both horizontal equity (population-based subareas) and vertical equity (bus-commuter-based subareas).}
\label{fig15}
\end{figure*} 

In Figure \ref{fig15}, the locations of charging stations and the corresponding number of charging piles in each station are depicted for both the population-based and bus-commuter-based merged subareas. The visualization considers fairness levels spanning from $0$ to $0.99$. It is worth noting that when $\eta = 0$, subplots (1) and (5) exhibit identical patterns. This similarity arises because the fairness constraint \eqref{cons:fair_socp} is not considered, resulting in planning results that remain consistent across different subareas and prioritize economic efficiency as the primary objective.

As the fairness level increases, variations become apparent between the results obtained from the population-based and bus-commuter-based merged subareas. These differences indicate that the two census-tract merging criteria capture distinct facets of transit equity and have some influence on the placement of charging infrastructure. However, it is important to note that the majority of charging station locations remain consistent across the population-based and bus-commuter-based merged subareas. For instance, when considering a fairness level of $0.99$, subplots (9) and (12) illustrate six charging station locations that are identical: nodes 19, 21, 31, 50, 53, and 55. Furthermore, these shared locations feature an equal number of installed charging piles. This finding is significant for decision-makers as it suggests that prioritizing the construction of charging stations in these shared locations would effectively address both horizontal and vertical equity concerns.

\begin{figure}[!h]
\centering
\includegraphics[width=0.9\linewidth]{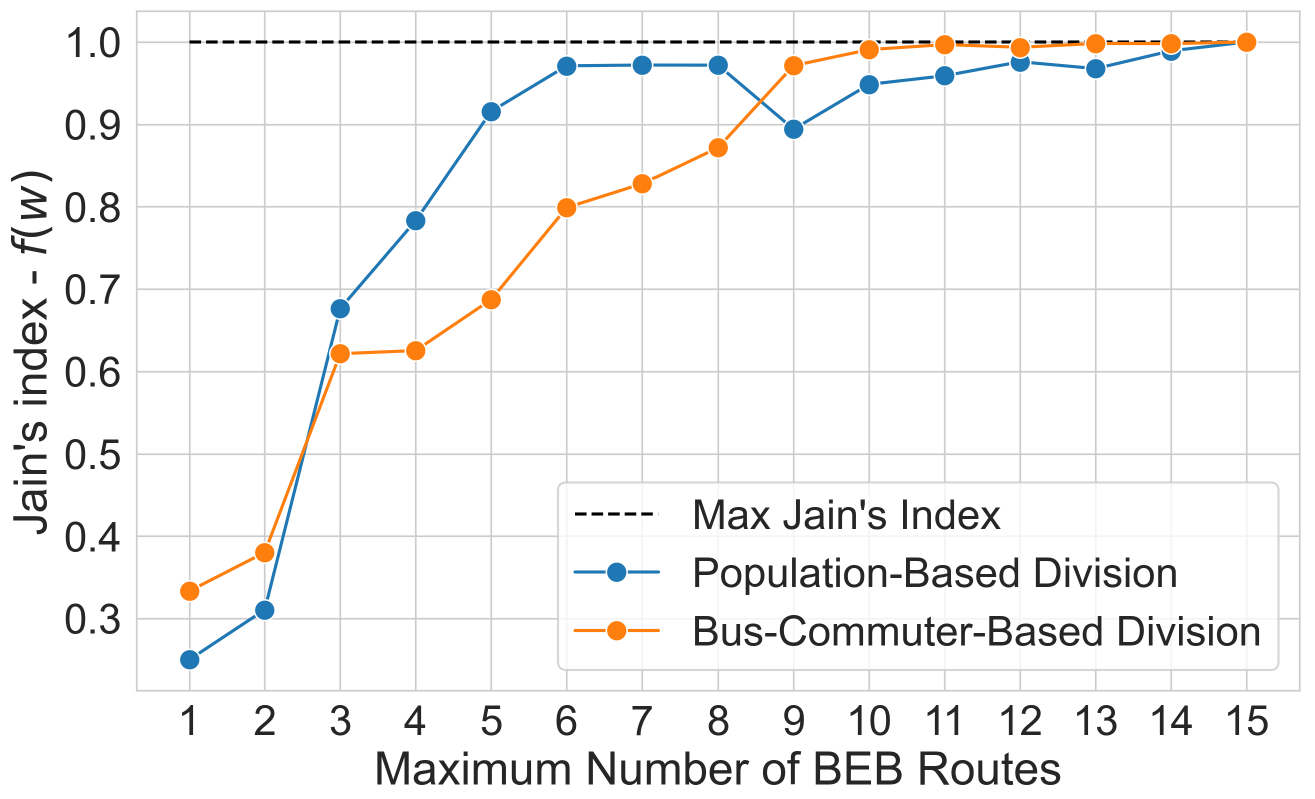}
\caption{Impact of varying maximum number of BEB routes on initial fairness level in planning results.}
\label{fig16}
\end{figure} 

The previous planning results with fairness consideration were obtained under the assumption that the budget could support up to five BEB routes. However, we have observed that the initial fairness index of the model is influenced by the choice of the maximum number of BEB routes to be invested. To visualize this relationship, we have created Figure \ref{fig16}, which illustrates the initial fairness index obtained without enforcing fairness constraint \eqref{cons:fair_socp} plotted against the maximum number of BEB routes. This graphical representation offers valuable insights into how the choice of the maximum number of BEB routes influences the fairness outcomes of the planning model when fairness is not explicitly considered.

The plot exhibits a general trend where the initial fairness index tends to increase as the maximum number of BEB routes increases. However, this trend does not strictly follow a monotonic increase within both the population-based and bus-commuter-based merged subareas. Notably, there is a decrease in the initial fairness index when the maximum BEB routes are set to 9 in the population-based merged subareas and 12 in the bus-commuter-based merged subareas. This finding suggests that even if there is more budget available to build charging infrastructure for BEBs, ignoring fairness constraints and pursuing the most economical planning results can lead to less transit equity. Therefore, it is crucial to account for fairness limitations when implementing BEB planning models.

\section{Conclusion}\label{sec6}
In this paper, a coupled power and transportation network framework is established for the planning of on-route charging infrastructure for BEBs. By integrating charging stations into both networks, we consider not only the investment cost of charging stations and charging piles but also additional investment in power lines and increased power loss costs in the power grid. These costs are minimized through the utilization of MISOCP. Additionally, we introduce fairness measurements into the planning results using Jain's index, which aligns well with the MISOCP model. This allows decision-makers to customize the level of fairness implemented during different phases of fleet electrification. All experiments in this study were conducted in South King County, a region recognized for being at the forefront of full electrification efforts. This area has been significantly impacted by air pollution, making it a pertinent location for our research.

Without fairness measurements, we compare the planning results under different levels of battery SOC when BEBs depart from origin stations. This analysis assists decision-makers in predicting the need for additional on-route charging infrastructure based on the current on-base charging station condition. Our siting and sizing results indicate that, regardless of the initial SOC of BEB batteries,  on-route charging stations are more likely to be located at stops serving multiple routes.

Furthermore, we incorporate a fairness measurement by imposing the fairness constraint in the planning model. By merging census tracts that intersect with bus routes into distinct subareas based on two tract features - the resident population and the population of bus commuters - we are able to measure both horizontal and vertical equity in the planning results. Comparing the planning outcomes under different fairness levels, we observe that a greater emphasis on fairness in the distribution of BEB routes among subareas results in higher planning costs. This information offers valuable insights to decision-makers on how to strike a balance between equity and economic efficiency in fleet electrification planning.

Our framework, which leverages the existing bus route map to create a virtual power network, has the potential to be applied to transportation systems in other cities. Additionally, our MISOCP model and fairness measurements provide practical guidance for allocating budgets and promoting social justice during the step-by-step electrification of bus fleets. In future research, we aim to extend the application of this planning model to larger transit systems and investigate acceleration algorithms to enhance its computational efficiency.

\section*{Acknowledgments}
This work is supported by the USDOT Tier 1 University Transportation Center TOMNET and National Science Foundation CMMI \#2053373. We would like to extend our special thanks to Dr. Cynthia Chen, Professor at the University of Washington, for her invaluable suggestions throughout this research project.

\printcredits

\newpage
\bibliography{scholar}

\end{sloppypar}
\end{document}